\documentclass[a4paper,11pt]{article}

\usepackage{color}
\usepackage{ulem}
\usepackage{hyperref}
\usepackage{enumerate}
\usepackage{amsmath}
\usepackage{graphicx}
\usepackage{float}
\usepackage{subfigure}

\def \be {\begin{equation}}
\def \ee {\end{equation}}
\def \ba {\begin{array}}
\def \ea {\end{array}}
\def \bea {\begin{eqnarray}}
\def \eea {\end{eqnarray}}

\def \ble {\begin{widetext}\begin{equation}}
\def \ele {\end{equation}\end{widetext}}
\def \blea {\begin{widetext}\begin{eqnarray}}
\def \elea {\end{eqnarray}\end{widetext}}

\def \nn {\nonumber}

\def \blea {\begin{widetext}\begin{eqnarray}}
\def \elea {\end{eqnarray}\end{widetext}}

\def \p {\partial}
\def \mP {\mathcal{P}} 
\def \mL {\mathcal{L}}

\begin{document}

\title{Parameter dependence of entanglement spectra in quantum field theories}


\author{Wu-zhong Guo\footnote{wuzhong@hust.edu.cn}, Jin Xu\footnote{2642467466@qq.com}}

\date{}
\maketitle

\vspace{-10mm}
\begin{center}
{\it School of Physics, Huazhong University of Science and Technology,\\
 Wuhan, Hubei
430074, China
\vspace{1mm}
}
\vspace{10mm}
\end{center}

\begin{abstract}
In this paper, we explore the characteristics of reduced density matrix spectra in quantum field theories. Previous studies mainly focus on the function $\mathcal{P}(\lambda):=\sum_i \delta(\lambda-\lambda_i)$, where $\lambda_i$ denote the eigenvalues of the reduced density matirx.  We introduce a series of functions designed to capture the parameter dependencies of these spectra. These functions encompass information regarding the derivatives of eigenvalues concerning the parameters, notably including the function $\mathcal{P}_{\alpha_J}(\lambda):=\sum_i \frac{\partial \lambda_i }{\partial \alpha_J}\delta(\lambda-\lambda_i)$, where $\alpha_J$ denotes the specific parameter. Computation of these functions is achievable through the utilization of R\'enyi entropy. Intriguingly, we uncover compelling relationships among these functions and demonstrate their utility in constructing the eigenvalues of reduced density matrices for select cases. We perform computations of these functions across several illustrative examples. Specially, we conducted a detailed study of the variations of $\mathcal{P}(\lambda)$ and $\mathcal{P}_{\alpha_J}(\lambda)$ under general perturbation, elucidating their physical implications. In the context of holographic theory, we ascertain that the zero point of the function $\mathcal{P}_{\alpha_J}(\lambda)$ possesses universality, determined as $\lambda_0=e^{-S}$, where $S$ denotes the entanglement entropy of the reduced density matrix. Furthermore, we exhibit potential applications of these functions in analyzing the properties of entanglement entropy.
\end{abstract}
\begin{center}
\tableofcontents
\end{center}

\maketitle

\section{Introduction}

Entanglement has emerged as a novel tool for discerning the structure of quantum field theories (QFTs) in recent years. Typically, various measures are introduced to quantify entanglement, with one of the most extensively studied being the entanglement entropy (EE). In certain QFTs, the entanglement entropy can be computed either analytically or numerically\cite{Holzhey:1994we,Srednicki:1993im,Vidal:2002rm,Calabrese:2004eu,Calabrese:2009qy,Casini:2009sr}. EE characterizes the quantum correlations between different types of spatial regions within field theory. Interestingly, within the framework of AdS/CFT\cite{Maldacena:1997re,Gubser:1998bc,Witten:1998qj}, entanglement entropy has been found to be related to minimal surfaces in the dual spacetime, following a law similar to the area law observed in black holes\cite{Ryu:2006bv,Hubeny:2007xt}. 

By partitioning the entire system into two parts, denoted as $A$ and its complementary $\bar A$, one can introduce the reduced density matrix $\rho_A := \text{tr}_{\bar A} \rho$, where $\rho$ represents the density matrix of the system. EE can then be considered as a function of $\rho_A$, defined as the Von Neumann entropy $S(\rho_A):=-tr \rho_A\log \rho_A$. In QFTs, the replica method via Euclidean path integrals is commonly employed to evaluate EE. It is necessary to initially compute the R\'enyi entropy, defined as
\begin{equation}
S^{(n)}(\rho_A) := \frac{\log \text{tr} \rho_A^n}{1-n},
\end{equation}
for a positive integer $n$. After analytically continuing $n$ to complex numbers, the entanglement entropy is expressed as $S(\rho_A) = \lim_{n \to 1} S^{(n)}(\rho_A)$. 

In QFTs, the trace in $\rho_A = \text{tr}_{\bar A} \rho$ is typically considered a formal definition. Unlike in finite-dimensional examples, obtaining $\rho_A$ directly through the trace operation seems unfeasible. Nevertheless, it's apparent that $\rho_A$ encompasses the complete information of the subsystem $A$. Consequently, reconstructing $\rho_A$ using entanglement measures becomes a significant area of investigation.  

The spectra of $\rho_A$ is studied in many-body system as a new topological order\cite{Li:2008kda}. 
In 2-dimensional conformal field theories (CFTs), the entanglement spectra can also be obtained using R\'enyi entropy \cite{Calabrese2008}, see also \cite{Hung:2011nu}. In \cite{Guo:2020rwj,Guo:2020roc}, the authors further investigate the entanglement spectra for the theory with holographic dual. An interesting result is that there exists an approximated state for any given states with holographic dual. Using the spectra decomposition, it is also possible to construct the so-called fixed area states in CFTs\cite{Guo:2021tzs}. The approximated state, as we mentioned above, can be understood as one special fixed area state.  The fixed area states are introduced in \cite{Akers:2018fow,Dong:2018seb} motivated by the similarity between AdS/CFT and quantum error correction (QEC) code\cite{Almheiri:2014lwa}. Hence, the entanglement spectra of $\rho_A$ holds significance in comprehending the entanglement structure of quantum field theories (QFTs), alongside elucidating the relationship between geometry and entanglement. There are also many studies on entanglement spectra in various directions, see ,e.g.,\cite{Yang,Ruggiero:2016yjt,Cho,Kudler-Flam:2021efr,Yan:2021yzy,Bai:2022obp}.

Generally, the R\'enyi entropy $S^{(n)}(\rho_A)$ encapsulates information about the spectra present within the reduced density matrix $\rho_A$. Numerous studies have investigated methods to obtain the density of the spectrum from R\'enyi entropy. For a specific theory and subsystem $A$, $S^{(n)}(\rho_A)$ is anticipated to depend on dimensional or dimensionless parameters, such as the subsystem's size, time, coupling constants of the theory. It is presumed that the spectra would be related to these parameters. Nevertheless, the density of the eigenvalues may not adequately capture the parameter-dependent nature of the spectrum. 

In this paper, we introduce a series of functions designed specifically to accomplish this objective, see details of the definitions in section.\ref{sectiongeneral}. Roughly, the density of eigenvalues represents the probability distribution of the eigenvalues. The functions presented in this paper aim to capture the changes in eigenvalues concerning a specific parameter.  For example, we introduce the function
\bea
\mP_{\alpha_J}(\lambda):=\sum_{i} \frac{\p \lambda_i}{\p \alpha_J}\delta(\lambda_i -\lambda),
\eea
where $\alpha_J$ is any parameter. The function $\mP_{\alpha_J}$ can be taken as the average value of $\frac{\p \lambda_i}{\p \alpha_J}$ at the eigenvalue $\lambda$.
These functions can be computed using the R\'enyi entropy, enabling an examination of the eigenvalue variations. Additionally, we uncover intriguing relationships among these functions.If the eigenvalues of $\rho_A$ satisfy more conditions, we can demonstrate the possibility of reconstructing the form of the eigenvalues using the results obtained from these functions. This has been carried out for a single interval in the vacuum state of 2-dimensional CFTs. The eigenvalues obtained by our method are consistent with the known results.

We have calculated these functions in several examples within 2-dimensional Conformal Field Theories (CFTs), including scenarios such as the single interval in the vacuum state, short intervals in arbitrary states, and obtaining a general result for the perturbation state 
$\rho+\delta\rho$. Based on these results, we have discussed the inherent properties of these functions. Additionally, we have made interesting observations regarding theories with a holographic dual. In the semi-classical limit, it has been found that the zero point of 
$\mP_{\alpha_J}(\lambda)$ is given by $\lambda_0=e^{-S}$, where $S$ is the EE for $\rho_A$. This particular value also appears in \cite{Guo:2020roc} where an approximated state for $\rho_A$ is constructed within the semi-classical limit. However, the relationship between these two results remains unclear.

We also delve into the potential application of these functions in characterizing the phase transition of Entanglement Entropy (EE). It has been observed that the shape of the function 
$\mP_{\alpha_J}$ does indeed mirror the variations in EE concerning the parameter. Our paper merely establishes a framework for studying the parameter dependence of the entanglement spectra in QFTs. On this basis, there exist numerous intriguing questions worthy of exploration.

The remainder of the paper is organized as follows. Section \ref{sectiongeneral} introduces a series of functions, including $\mP$ and $\mP_{\alpha_J}$, which describe the entanglement spectra along with their parameter dependencies and discusses their properties. Following this, Section \ref{section_example} presents the calculation of $\mP$ and $\mP_{L}$ in the vacuum state of 2D CFTs as an illustrative example. In section \ref{section_arbitrary state} we delve into the calculation for an arbitrary state of 2D CFTs with a short interval. Notably, it reveals a shift in the zero point of $\mP_l$ compared to the vacuum state. Section \ref{section_perturbation} examines the scenario where the density matrix experiences a perturbation, denoted as $\rho=\rho_0+\delta\rho$. This section investigates the alterations in $\mP$ and $\mP_{\alpha_J}$ subsequent to the perturbation. We also provide explanations for each term in the obtained results.
In Section \ref{section_holographic}, the paper explores the computation of $\mP$ and $\mP_{\alpha_J}$ in holographic theory, employing the saddle point approximation. Furthermore, it discusses the zero point of $\mP_{\alpha_J}$ within this context and find a universal result of the zero point. Section \ref{section_7} extends the discussion to analyze the derivative of entanglement entropy using the function $\mP_{\alpha_J}$. Finally, Section \ref{section_conclusion} presents the concluding remarks. Detailed calculations are provided in the appendices.

\section{General setup}\label{sectiongeneral}
Assume the spectra of $\rho_A$ are $\{\lambda_i\}$. We can define the spectra density as
\bea\label{p0}
\mP(\lambda):= \sum_i \delta(\lambda_i -\lambda).
\eea 
Roughly, it can be understood as the number of degenerate eigenstates for the eigenvalue $\lambda$.\\
By the definition, it's easy to know that it has property
\begin{align} \label{p_property}
\int_{-\infty}^{+\infty}f(\lambda)\mP(\lambda)d\lambda
&=\sum_if(\lambda_i).
\end{align} 
For example, when $f(\lambda)=\lambda$, $\int_{-\infty}^{+\infty}\lambda \mP(\lambda)d\lambda=\sum_i\lambda_i=1$; when $f(\lambda)=-\lambda\log{\lambda}$, $\int_{-\infty}^{+\infty} -\lambda\log{\lambda} \mP(\lambda)d\lambda=-\sum_i\lambda_i\log{\lambda_i}=S_\mathcal{A}$. From the above example, it can also be seen that when we obtain $\mP(\lambda)$, the entanglement entropy $S_\mathcal{A}$ can be easily calculated, so it can be seen that the information of the entanglement spectra is greater than the entanglement entropy.\\
Notice in $\{\lambda_i\}$, $\lambda_i>0$ and there is a maximum eigenvalue $\lambda_{m}$, so we can rewrite (\ref{p_property}) to
\bea\label{p_property2}
\int_{0}^{\lambda_{m}}f(\lambda)P(\lambda)d\lambda=\sum_if(\lambda_i).
\eea 
Generally, the eigenvalue $\lambda_i$ can be taken as functions of some parameters denoted by $\{\alpha_J \}$. To characterize the relation we would like to introduce and explore a new quantity
\bea\label{p1}
\mP_{\alpha_J}(\lambda):= \sum_i \frac{\p \lambda_i}{\p \alpha_J}\delta(\lambda_i -\lambda).
\eea
The function $\bar \mP_{\alpha_J}(\lambda):=\frac{\mP_{\alpha_J}(\lambda)}{\mP(\lambda)}$ can be taken as the average value of $\frac{\p \lambda_i}{\p \alpha_J}$ for the eigenvalue $\lambda$. It is obvious that we should have the constraint 
\bea
\int_{0}^{\lambda_m} d\lambda \mP_{\alpha_J}(\lambda)=0,
\eea
where $\lambda_m$ is the maximal eigenvalue of $\rho_A$. The above constraint comes from the normalization of the reduced density matrix.
We also interest at
\bea\label{p2}
\mP_{\alpha_{J m}}:=  \sum_i \frac{\p^m \lambda_i}{\p \alpha_{J}^m}\delta(\lambda_i -\lambda),
\eea
\bea\label{p3}
\mP_{\alpha_J^m}:=\sum_i \left(\frac{\p \lambda_i}{\p \alpha_J}\right)^m\delta(\lambda_i -\lambda),
\eea
for the integer $m$.

Most generally, we can also define the following quantities,
\begin{align}
\mP_{(\alpha_{J_{11}}... \alpha_{J_{1m_1}})...(\alpha_{J_{n1}}... \alpha_{J_{n m_n}})}:=  \sum_i \frac{\p^{m_1} \lambda_i}{\p \alpha_{J_{11}}...\p {\alpha_{J_{1m_1}}}}...\frac{\p^{m_n} \lambda_i}{\p \alpha_{J_{n1}}...\p {\alpha_{J_{n m_n}}}}\delta(\lambda_i -\lambda),   
\end{align}
for given integer $n$ and $\{m_1,...m_n\}$. We can define the order $\mathcal{N}$ of the functions by counting the power of the derivatives, $\mathcal{N}:= \sum_{i=1}^{n}m_i$.

Specially, we can get
\bea\label{p4}
\mP_{\alpha_{J_1}... \alpha_{J_m}}:= \mP_{(\alpha_{J_1}... \alpha_{J_m})}= \sum_i \frac{\p^m \lambda_i}{\p \alpha_{J_1}...\p {\alpha_{J_m}}}\delta(\lambda_i -\lambda),
\eea
\bea\label{p5}
\mP_{(\alpha_{J_1})... (\alpha_{J_m})} = \sum_i \frac{\p \lambda_i}{\p \alpha_{J_1}}...\frac{\p \lambda_i}{\p \alpha_{J_m}}\delta(\lambda_i -\lambda),
\eea
for a given integer $m$. When $\alpha_{J_1}=\alpha_{J_2}=...=\alpha_{J_m}=\alpha_J$, (\ref{p4}) becomes (\ref{p2}), and (\ref{p5}) becomes (\ref{p3}).

By the definition the above functions 
 are determined once all the eigenvalues are given. But in most cases, specially examples in QFTs,  we have very limit information about the eigenvalues. Our motivation to define these functions is to gain more information on the distribution and parameter dependence of the eigenvalues. These functions appear to be independent, but we will demonstrate later that there are connections between them, which are implicit in their definitions.

\subsection{Relations among the functions}

All the functions that we defined in last section can be evaluated by the R\'enyi entropy. Recall the definition of R\'enyi entropy
\bea\label{renyi}
S^{(n)}=\frac{1}{1-n} \log{{\rm Tr}_{\mathcal{A}} \rho_\mathcal{A}^n}=\frac{1}{1-n} \log{\sum_i\lambda_i^n}.
\eea
By using the property (\ref{p_property2}), the above equation can be rewritten as
\begin{align}\label{renyi_1}
\sum_i\lambda_i^n&=e^{(1-n)S^{(n)}}\nn\\
\sum_i \int_{0}^{\lambda_{m}} \lambda^n \delta(\lambda_i-\lambda) d\lambda&=e^{(1-n)S^{(n)}}\nn \\
\int_0^{\lambda_m} \lambda^{n}\mP(\lambda)d\lambda&=e^{(1-n)S^{(n)}}.
\end{align}
Compare the form of Laplace transformation:
\bea\label{laplace}
\mathcal{L}\left[f(t)\right]:=\int_{0}^{\infty} e^{nt} f(t) dt.
\eea
We find that the above formula (\ref{renyi_1}) is similar to the form of Laplace transformation (\ref{laplace}). For further calculation, let $\lambda=e^{-b-t}$, where $b=-\log{\lambda_m}$.\\
Actually, from (\ref{renyi}) we have $b=\lim_{n\to\infty}S^{(n)}$.
By using the above replace, it's easy to know $\lambda =0$ corresponding $t=\infty$ and $\lambda = \lambda_m $ corresponding $t = 0 $, so (\ref{renyi_1}) becomes 
\begin{align}
\mathcal{L}\left[\mP(e^{-b-t} )e^{-b-t}\right]=e^{(1-n)S^{(n)}}e^{nb}.
\end{align}
By using inverse Laplace transformation, we can obtain the density of eigenvalue $\mP$
\bea\label{Inverse10}
\mP(e^{-b-t})= \lambda^{-1} \mL^{-1}\left[e^{nb+(1-n)S^{(n)}}\right],
\eea
where $\mL^{-1}$ is defined as
\bea
f(t)=\mL^{-1}\left[F(n)\right]=\frac{1}{2\pi i} \int_{\gamma_0-i\infty}^{\gamma_0+i\infty} F(n) e^{nt} dn,
\eea
$\gamma_0$ is chosen for the convergence of the integration. 

So we can say that density of eigenvalue $\mP$ and R\'enyi entropy $S^{(n)}$ are each other (inverse) Laplace transformation.

If taking derivative with respect to $\alpha_J$ for both side of the above equation(\ref{renyi}), we have
\bea\label{d1}
\sum_i n\lambda_i^{n-1} \frac{\p \lambda_i }{\p \alpha_J}=(1-n)\frac{\p S^{(n)}}{\p \alpha_J}e^{(1-n)S^{(n)}}.
\eea
By using the definition (\ref{p1}), the above equation can be rewritten as
\bea
\int_0^{\lambda_m}d\lambda\  n \lambda^{n-1}\mP_{\alpha_J}(\lambda)=(1-n)\frac{\p S^{(n)}}{\p \alpha_J}e^{(1-n)S^{(n)}}.
\eea
Let $\lambda=e^{-b-t}$, where $b=-\log{\lambda_m}$, we have
\bea
\int_0^\infty dt e^{-n t} \mP_{\alpha_J}(e^{-b-t})=\frac{1-n}{n}e^{nb+(1-n)S^{(n)}}\frac{\p S^{(n)}}{\p \alpha_J}.
\eea
Similar as the case for density of eigenvalue $\mP$(\ref{Inverse10}), one could evaluate $\mP_{\alpha_J}$ by using inverse Laplace transformation, that is
\bea\label{Inverse1}
\mP_{\alpha_J}(e^{-b-t})= \mL^{-1}\left[\frac{1-n}{n}e^{nb+(1-n)S^{(n)}}\frac{\p S^{(n)}}{\p \alpha_J}\right].
\eea
Note that the expression in the square bracket is a function of $n$. One could obtain $\mP$ and $\mP_{\alpha_J}$ once knowing  the R\'enyi entropy $S^{(n)}$. 

By using the property of inverse Laplace transformtion, one could derive the relations between the functions. Using (\ref{Inverse10}),(\ref{Inverse1}) we find
\bea\label{r1}
\frac{\p \mP(\lambda)}{\p \alpha_J}=-\frac{\p \mP_{\alpha_J}(\lambda)}{\p \lambda}.
\eea

Futher taking derivative with respect to $a_J$ for both side of (\ref{d1}),  other quantities (\ref{p2}) and (\ref{p3})  would appear. One could obtain these quantities by similar method as above. For example, taking twice derivative we would obtain
\begin{align}
\sum_i n(n-1)\lambda_i^{(n-2)}(\frac{\partial \lambda_i}{\partial \alpha_J})^2+\sum_i n\lambda_i^{(n-1)}\frac{\partial^2 \lambda_i}{\partial \alpha_J^2}=\frac{\p ^2}{\p L^2} e^{(1-n)S^{(n)}},    
\end{align}
Similarly, we will have
\bea 
\int_{0}^{\lambda_{m}} n(n-1)\lambda^{n-2} P_{\alpha_J^2}(\lambda) d\lambda +\int_{0}^{\lambda_{m}} n\lambda^{n-1} P_{\alpha_{J 2}}(\lambda) d\lambda =\frac{\partial^2}{\partial \alpha_J^2} e^{(1-n)S^{(n)}}.
\eea
By using the property of inverse Laplace transformation and (\ref{Inverse1}), we find
\bea\label{r2}
\frac{\p \mP_{\alpha_J}}{\p \alpha_J}=\mP_{\alpha_{J 2}}-\frac{\p \mP_{\alpha_J^2}}{\p \lambda}.
\eea
For higher power we can also obtain similar relations as we will show in next section. 

\subsection{Consistent with definition}\label{consistentsection}
In fact the relations of the functions are also consistent with the definition of these functions. By the definition of $\mP$, taking derivative with respect to $\alpha_J$  for $\mP$ we have
\bea
\frac{\p \mP}{\p \alpha_J}=\sum_i \frac{\p }{\p \alpha_J} \delta(\lambda_i-\lambda)=\sum_i \frac{\p \lambda_i}{\p \alpha_J}\delta'(\lambda_i-\lambda).
\eea 
$\mP$ depends on the parameter $\alpha_J$ through $\lambda_i$. One should keep in mind that $\lambda$ is independent with the parameter. Similarly, $\frac{\p \lambda_i}{\p \alpha_J}$ is also independent with $\lambda$, thus we find
\bea\label{consistent1}
\frac{\p \mP}{\p \alpha_J}=-\frac{\p}{\p \lambda}\sum_i \frac{\p \lambda_i}{\p \alpha_J}\delta(\lambda_i-\lambda)=-\frac{\p \mP_{\alpha_J}}{\p \lambda}.
\eea 
By the same logic, we can derive the relation (\ref{r2}) as
\begin{align}
\frac{\partial }{\partial \alpha_J} P_{\alpha_J}(\lambda) &= \frac{\partial }{\partial \alpha_J} \sum_i \frac{\partial \lambda_i}{\partial \alpha_J} \delta(\lambda_i-\lambda) \nn \\
&= \sum_i \frac{\partial^2 \lambda_i}{\partial \alpha_J^2} \delta(\lambda_i-\lambda) +\sum_i \frac{\partial \lambda_i}{\partial \alpha_J} \frac{\partial \lambda_i}{\partial \alpha_J}\delta'(\lambda_i-\lambda) \nn \\
&= \sum_i \frac{\partial^2 \lambda_i}{\partial \alpha_J^2} \delta(\lambda_i-\lambda) +\sum_i (\frac{\partial \lambda_i}{\partial \alpha_J})^2 \times \left(-\frac{\partial }{\partial \lambda} \delta(\lambda_i-\lambda) \right)\nn \\
&=P_{\alpha_{J2}}(\lambda)-\frac{\partial }{\partial \lambda} P_{\alpha_J^2}(\lambda).    
\end{align}
Most generally, we can get that
\begin{align}\label{generally}
&\frac{\partial }{\partial \alpha_K} \mP_{(\alpha_{J_{11}}... \alpha_{J_{1m_1}})(\alpha_{J_{21}}... \alpha_{J_{2m_2}})...(\alpha_{J_{n1}}... \alpha_{J_{n m_n}})} \nn \\
=&\mP_{(\alpha_{J_{11}}... \alpha_{J_{1m_1}} \alpha_K)(\alpha_{J_{21}}... \alpha_{J_{2m_2}})...(\alpha_{J_{n1}}... \alpha_{J_{n m_n}})} \nn \\
&+\mP_{(\alpha_{J_{11}}... \alpha_{J_{1m_1}} )(\alpha_{J_{21}}... \alpha_{J_{2m_2}}\alpha_K)...(\alpha_{J_{n1}}... \alpha_{J_{n m_n}})} \nn \\
&+...... \nn \\
&+\mP_{(\alpha_{J_{11}}... \alpha_{J_{1m_1}} )(\alpha_{J_{21}}... \alpha_{J_{2m_2}})...(\alpha_{J_{n1}}... \alpha_{J_{n m_n}}\alpha_K)} \nn \\
&-\frac{\partial }{\partial \lambda} \mP_{(\alpha_{J_{11}}... \alpha_{J_{1m_1}} )(\alpha_{J_{21}}... \alpha_{J_{2m_2}})...(\alpha_{J_{n1}}... \alpha_{J_{n m_n}})(\alpha_K)}.    
\end{align}
Please see the Appendix.\ref{prove1} for more details of the calculations. 

With relation (\ref{generally}), it is easy to get an interesting conclusion about two parameters. Consider two unrelated parameters $\alpha_{J_{1}}$ and $\alpha_{J_{2}}$, it's easy to get
\begin{align}\label{two_parameters}
\frac{\partial }{\partial \alpha_2} \mP_{\alpha_{J_{1}}}
&=\frac{\partial }{\partial \alpha_1} \mP_{\alpha_{J_{2}}}.
\end{align}

\subsection{With further assumptions}\label{sectionfurther}
In the above discussions we find the functions at order $\mathcal{N}$ would have some relations. For example, for $\mathcal{N}=2$, $\mP_{\alpha_J^2}$ and $\mP_{\alpha_{J2}}$ are not independent. In fact this means one can not solve $\mP_{\alpha_J^2}$ and $\mP_{\alpha_{J2}}$ seperately by only using R\'enyi entropy. To obtain them we should have more assumptions.

In general, $\frac{\p \lambda_i}{\p \alpha_J}$ should not depend on the eigenvalue $\lambda_i$. But in some special case we find that $\frac{\p \lambda_i}{\p \alpha_J}$ can still be seen as a function of $\lambda_i$, that is $\frac{\p \lambda_i}{\p \alpha_J}=f(\lambda_i,\alpha_J)$. We do not expect this is true for general cases. In appendix we use simple examples to show this. For the special case we find all the functions can be solved. 

With the assumption we have
\bea\label{Ker}
\mP_{\alpha_J}(\lambda)=\sum_i f(\lambda_i,\alpha_J) \delta(\lambda_i -\lambda)=f(\lambda,\alpha_J)\sum_i \delta(\lambda_i -\lambda)=f(\lambda,\alpha_J) \mP(\lambda).
\eea 
By using (\ref{Ker}) we have the equation
\bea\label{diffequation}
\frac{\p \lambda_i}{\p \alpha_J}=f(\lambda_i,\alpha_J)=\frac{\mP_{\alpha_J}(\lambda_i)}{\mP(\lambda_i)}.
\eea
where $\mP$ and $\mP_{\alpha_J}$ can be obtained by $S^{(n)}$. Once knowing $S^{(n)}$, one could solve the equation with suitable conditions.  With these results one could obtain more details of the eigenvalues $\lambda_i$. By choosing more parameters $\alpha_J$ one could reconstruct the eigenvalues of $\rho_A$. In the following we will show some examples. On the contrary one may assume the eigenvalues $\lambda_i$ satisfy the relation $\frac{\p \lambda_i}{\p \alpha_J}=f(\lambda_i,\alpha_J)$. If the differential equation has no proper solutions, one can conclude this assumption is false. 

The higher order function can also be associated with $\mP$. For example, by definition
\bea\label{high1}
\mP_{\alpha_J^2}(\lambda)=f(\lambda,\alpha_J)^2\mP(\lambda),
\eea
\begin{align}\label{high2}
\mP_{\alpha_{J2}}(\lambda)&=\sum_i \frac{d }{d \alpha_J}f(\lambda_i,\alpha_J) \delta(\lambda_i-\lambda)\nn\\
&=\sum_i (\frac{\partial }{\partial \lambda_i}f(\lambda_i,\alpha_J) \frac{\partial \lambda_i}{\partial \alpha_J} + \frac{\partial }{\partial \alpha_J}f(\lambda_i,\alpha_J) )\delta(\lambda_i-\lambda)\nn\\
&=\sum_i (\frac{\partial f(\lambda_i,\alpha_J)}{\partial \lambda_i} f(\lambda_i,\alpha_J) + \frac{\partial }{\partial \alpha_J}f(\lambda_i,\alpha_J) )\delta(\lambda_i-\lambda)\nn\\
&=(\frac{\partial f(\lambda,\alpha_J)}{\partial \lambda} f(\lambda,\alpha_J) + \frac{\partial }{\partial \alpha_J}f(\lambda,\alpha_J) ) \sum_i \delta(\lambda_i-\lambda)\nn\\
&=(\frac{\partial f(\lambda,\alpha_J)}{\partial \lambda} f(\lambda,\alpha_J) + \frac{\partial }{\partial \alpha_J}f(\lambda,\alpha_J) ) \mP(\lambda).    
\end{align}
So we can write $\mP_{\alpha_J}(\lambda)$, $\mP_{\alpha_J^2}(\lambda)$ and $\mP_{\alpha_{J2}}(\lambda)$ just by $f(\lambda,\alpha_J)$ and $\mP(\lambda)$. We can test the self-consistency of (\ref{Ker}), (\ref{high1}) and (\ref{high2}) through the relation (\ref{r2}).

More generally, we want to write $\mP_{\alpha_J^m}(\lambda)$ and $\mP_{\alpha_{Jm}}(\lambda)$ by $f(\lambda,\alpha_J)$ and $\mP(\lambda)$. We define a new derivation
\bea
\frac{D}{D \alpha_J}=f(\lambda,\alpha_J) \frac{\partial}{\partial \lambda}  + \frac{\partial }{\partial \alpha_J}.
\eea
So $\mP_{\alpha_{J2}}(\lambda)$ can be rewritten as $\mP_{\alpha_{J2}}(\lambda)=\frac{D f(\lambda,\alpha_J)}{D \alpha_J} P(\lambda)$

One could show that
\bea\label{higher_times}
\mP_{\alpha_J^m}(\lambda)=f(\lambda,\alpha_J)^m \mP(\lambda),
\eea
\bea\label{higher_derivative}
\mP_{\alpha_{J m}}(\lambda)=\frac{D^{m-1}f(\lambda,\alpha_J)}{D \alpha_J^{m-1}} \mP(\lambda),
\eea
where $m\in Z$ and $m\geq2$. See the Appendix.\ref{prove2} for more details of the calculations.

\section{Examples in 2-dimensional CFTs}\label{section_example}
There are many known analytic results of R\'enyi entropy of one interval for some states in 2-dimensional CFTs. Using these results we could directly obtain the functions discussed in previous sections. We will firstly evaluate $\mathcal{P}$ and $\mathcal{P}_{\alpha_J}$ for $\alpha_J$ being the size of the interval $L$ and central charge $c$. With further assumption one could obtain the eigenvalues. The eigenvalues of this example can be derived by conformal mapping method used in \cite{Cardy:2016fqc}. Our results are consistent with the ones in \cite{Cardy:2016fqc}.  
\subsection{One interval on infinite line, vacuum state}\label{examplevacuumsection}
Using replica method, we can get the R\'enyi entropy $S^{(n)}(\rho_A)$ of one interval on the infinite line in the vacuum state of 2D CFTs\cite{Calabrese:2004eu}\cite{Calabrese:2009qy}
\bea\label{vacuum renyi}
S^{(n)}=\frac{c}{6}(1+\frac{1}{n})\log{\frac{L}{\epsilon}},
\eea
where $L$ is the length of system $A$, $\epsilon$ is the UV cut-off. 
We have $b=-\log{\lambda_m}=\frac{c}{6}\log{L/\epsilon}$.

Using (\ref{Inverse10}) and (\ref{Inverse1}), we can get
\bea\label{spectra1}
&&\mP(\lambda)=\frac{1}{\lambda}\left[\frac{\sqrt{b} I_1\left(2 \sqrt{bt}\right)}{\sqrt{t}}+\delta (t)\right],\nn \\
&&\mP_L(\lambda)=-\frac{c}{6L}\left[\frac{(b-t) I_1\left(2 \sqrt{bt}\right)}{\sqrt{b t}}+\delta (t)\right],\nn \\
&&\mP_c(\lambda)=-\frac{\log{L}}{6}\left[\frac{(b-t) I_1\left(2 \sqrt{bt}\right)}{\sqrt{b t}}+\delta (t)\right],
\eea
where $I_n(x)$ is the modified Bessel functions of the first kind, we have the relation $\lambda=e^{-b-t}$.
\begin{figure}
\centering
\subfigure{\includegraphics[scale=0.48]{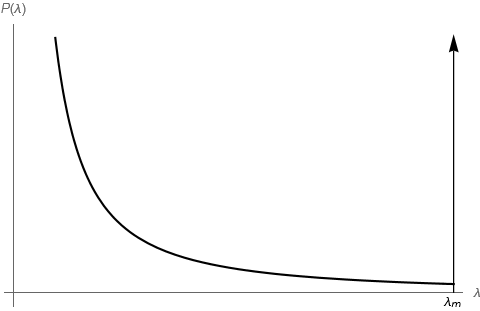}}
\subfigure{\includegraphics[scale=0.48]{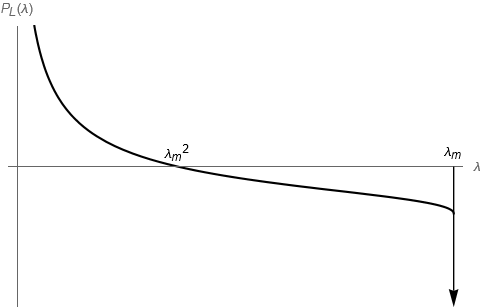}}
\caption{The illustration of $\mP(\lambda)$ and $\mP_L(\lambda)$ in the case of the vacuum state of 2D CFT, where the arrows represent the Dirac delta function.}
\label{plot_P}
\end{figure}
It is straightforward to check that (\ref{spectra1})  satisfy the relation (\ref{r1}), that is 
\bea
\frac{\partial}{\partial L}  P(\lambda)=-\frac{\partial}{\partial \lambda}  P_L(\lambda).
\eea
See the Appendix.\ref{sectionconsistent} for the details. We also find the relation (\ref{two_parameters}),
\bea
\frac{\partial}{\partial c}  P_L(\lambda)=\frac{\partial}{\partial L}  P_c(\lambda)
\eea
See the Appendix.\ref{sectionconsistent} for the details.\\
$\mathcal{P}_L(\lambda)$ as a function of $\lambda$ can be used to reflect how the eigenvalues change with the scale of the subsystem. For the maximal eigenvalue $\lambda_m$, one could obtain $\frac{\p \lambda_m}{\p L}$ by using $-\log{\lambda_m}=\frac{c}{6}\log{L/\epsilon}$.  One could also check that $\frac{\p \lambda_m}{\p L}=\tilde \mP(\lambda_m)$. For $\lambda \ne \lambda_m$, there is a zero point of the function $\mP_L$, which is given by $t=b$ or $\lambda_0=e^{-2b}=\lambda_m^2$. For $\lambda< \lambda_0$, $\mP_L(\lambda)>0$, which means that as the scale of the subsystem increases, the average eigenvalues smaller than $\lambda_0$ are increasing. While for $\lambda>\lambda_0$, $\mP_L(\lambda)<0$, the average eigenvalues are decreasing. The physical significance of the zero point is not very clear, but we can see that the function $\mP_L(\lambda)$ must have at least one zero because the integral result of it should be zero. We plot the function $\mP(\lambda)$ and $\mP_L(\lambda)$ in Fig.\ref{plot_P}.

\subsection{One interval on cylinder, vacuum state}\label{sectioncylinder}
Consider the CFT is defined on a cylinder with circumference $R$. The interval is $A=[0,L]$ with length $L$. The R\'enyi entropy for this case is given by
\bea
S^{(n)}=\frac{c}{6}(1+\frac{1}{n})\log \left(\frac{R}{\epsilon\pi}\sin \frac{\pi L}{R}\right).
\eea
We have $b=-\log \lambda_m=\frac{c}{6}\log \left(\frac{R}{\epsilon\pi}\sin \frac{\pi L}{R}\right)$. It is straightforward to obtain the functions
\bea
&&\mP(\lambda)=\frac{1}{\lambda}\left[\frac{\sqrt{b} I_1\left(2 \sqrt{bt}\right)}{\sqrt{t}}+\delta (t)\right],\nn \\
&&\mP_L(\lambda)=-\frac{c}{6}\frac{\pi  \cot \left(\frac{\pi  L}{R}\right)}{R}\left[\frac{(b-t) I_1\left(2 \sqrt{b t}\right)}{\sqrt{b t}}+\delta (t)\right].
\eea
One could check that $\int_{0}^{\lambda_m} \lambda \mP(\lambda)d\lambda=1$ and $\int_{0}^{\lambda_m} \mP_L(\lambda)d\lambda=0$. 

\begin{figure}
\centering
\includegraphics[scale=0.47]{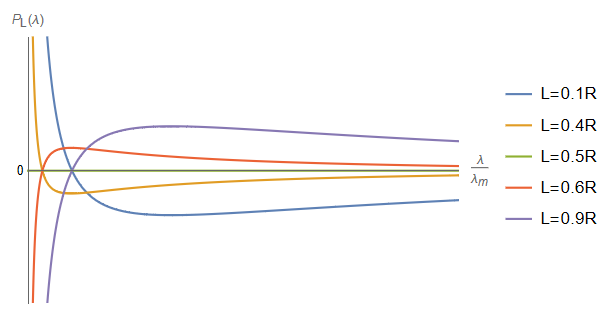}
\caption{The plots of $\mP_L(\lambda)$ with various parameters in the scenario of a single interval on a cylinder. We omit the representation of the term $\delta(\lambda_m-\lambda)$ in the plot as it isn't important for our current discussion.}
\label{plot_PL_R}
\end{figure}

For $P_L(\lambda)$ we also have a zero point at $t_0=b$ or $\lambda_0 =e^{-2b}=\lambda_m^2$. The figure of $\mP_L(\lambda)$ is slightly different from the case on infinite line. If $L<R/2$, we have $\mP_{L}>0$ for $0<\lambda<\lambda_0$, while $\mP_L<0$ for $\lambda_0<\lambda<\lambda_m$. The result is similar as the case on infinite line. But for $L>R/2$, the figure is flipped.  We have $\mP_{L}<0$ for $0<\lambda<\lambda_0$, while $\mP_L>0$ for $\lambda_0<\lambda<\lambda_m$. There exists a critical point $L=R/2$, the function $\mP_L$ is vanishing at this point. We show the results in Fig.\ref{plot_PL_R}.  

\subsection{Reconstruction of the eigenvalues with further assumption}\label{reconstructionsection}
Without further assumption one can not obtain higher order functions as we have shown in previous section. For the present example we find the maximal eigenvalue satisfies $\frac{\p \lambda_m}{\p L}=\bar \mP(\lambda_m)$. Let us assume for other eigenvalues we also have $\frac{\p \lambda_i}{\p L}=f(\lambda_i,L)$. From the discussions in section.\ref{sectionfurther}, using 
 (\ref{spectra1}), we have
\bea\label{diffferential}
\frac{\partial \lambda_i}{\partial L}=\frac{-\log{\lambda_i}-2b}{b}\frac{c}{6L}\lambda_i.
\eea
Solving the above differential equation, we can get
\bea\label{lamda1}
\lambda_i=e^{-\frac{C_i}{\log{L/\epsilon}}-b},
\eea
where $C_i$ are constants unrelated to $L$.

\subsection{Eigenvalues of modular Hamiltonian by conformal mapping}\label{sectionconformalmap}
For the one interval example in last sections, one could reconstruct the eigenvalues of the modular Hamiltonian by using the functions $\mathcal{P}_L$ and $\mathcal{P}$. For the simple example one could obtain the eigenvalues by the methods explored in \cite{Cardy:2016fqc}. As the R\'enyi entropy and entanglement entropy are UV divergence, we should introduce some regulator to obtain the results. Let us focus on the 2-dimensional CFTs. In \cite{Cardy:2016fqc} the authors show one could only consider the states that are projected out the basis in a small spatial region of thickness $\epsilon$ around the common boundary of $A$ and $\bar A$. In the Euclidean path integral representation of the reduced density matrix $\rho_A$, this is to introduce a hole around the boundary point of $A$. Some suitable boundary conditions should be imposed on the boundary of the hole. For the one interval example, the topology of the manifold is an annulus. 

Suppose $A=[0,L]$. The system is in the vacuum state on the infinite line.   The corresponding state is associated with the Euclidean spacetime with a disk of radius $\epsilon$ removed at endpoints  of $A$, which can be mapped to the annulus by the conformal mapping,
\bea\label{conformalmap}
w=\log \frac{z}{L-z},
\eea
where $w$ is the coordinate of the annulus.  The width of the annulus is $W=f(L-\epsilon)-f(\epsilon)\simeq 2\log \frac{L}{\epsilon}=\frac{12b}{c}$. The modular Hamiltonian $K_A$ is locally generator of rotation around the endpoints of $A$. Under the conformal map (\ref{conformalmap}) $K_A$ is mapped to the time evolution operator $H_w:=\int dv T_{vv}$ along the direction $v:=Im (w)$ upto some constants, $K_A$ and $H_w$ are unitarily equivalent. Thus the eigenvalues of $K_A$ should be same as $H_w$. By using $T_{vv}=T(w)+\bar T(\bar w)$ we have
\bea
H_w=\int dw T(w) +\int d\bar w \bar T(\bar w).
\eea
Under the conformal transformation (\ref{conformalmap}), we obtain
\bea\label{HwKA}
H_w=K_A+\frac{c}{12} \int_{\epsilon}^{L-\epsilon}dx \frac{L}{x(L-x)}=K_A+b,
\eea
where the constant term is from the Schwartzian term and we define 
\bea
K_A:=\int_{\epsilon}^{L-\epsilon} dz \frac{z(L-z)}{L}T(z)+\int_{\epsilon}^{L-\epsilon} d\bar z \frac{\bar  z(L-\bar z)}{L}\bar T(\bar z),
\eea
which is the regularized modular Hamiltonian of the single interval on infinite line. 

For CFTs on the annulus with width $W$ the eigenvalues of $H_w$ are given by $\frac{\Delta_i -\frac{c}{24}}{W}$. Thus by using (\ref{HwKA}) the eigenvalues of $K_A$ is given by  $\frac{\Delta_i -\frac{c}{24}}{W}-b$. In paper \cite{Guo:2020roc} the author shows the reduced density matrix $\rho_A=e^{-K_A-2b}$ by normalization. Therefore we expect the eigenvalues of $\rho_A$ should be $e^{-\frac{\Delta_i -\frac{c}{24}}{W}-b}$, which is just the form as (\ref{lamda1}).

\section{Short interval in arbitrary state}\label{section_arbitrary state}
Computing the Renyi entropy of an arbitrary state is usually very challenging, but in some cases, we can obtain the result perturbatively using the Operator Product Expansion (OPE) of the twistor operator\cite{Cardy:2007mb,Headrick:2010zt,Calabrese:2010he,Rajabpour:2011pt,Chen:2013kpa}. In this section we will focus on 2-dimensional CFT with a short interval. 
\subsection{R\'enyi entropy for arbitrary state}
Assume the length of the interval is $l$ and the state is $\rho$. For simplicity, we only list the contributions from the operators in the vacuum conformal family, such as $T$, $\bar T$, and $\mathcal{A}$, and assume the state is translationally invariant.  Up to $O(l^4)$ the R\'enyi entropy can be expanded in term of $l$ as\footnote{The extension of the calculation to arbitrary situations and higher order is straightforward. In the following we will use this result for holographic CFTs, in which the contributions from the vacuum conformal family is dominant.}
\begin{align}\label{Renyishort}
S^{(n)}=&\frac{c}{6}\frac{n+1}{n}\log{\frac{l}{a}}+\frac{n+1}{n}k_2 l^2\nn\\
&+[\frac{(n+1)(n^2-1)}{n^3}k_4+\frac{(n+1)(n^2+11)}{n^3}k'_4]l^4+O(l^6),
\end{align}
with
\bea
&&k_2=-\frac{1}{12}\left(\langle T \rangle_\rho+\langle \overline{T} \rangle_\rho\right),\nn\\
&&k_4=-\frac{1}{288}\left(\langle A \rangle_\rho+\langle \overline{A} \rangle_\rho\right)+\frac{1}{288}\left(\langle T \rangle^2_\rho+\langle \overline{T} \rangle^2_\rho\right),\nn\\
&&k'_4=-\frac{1}{720c}\left(\langle T \rangle^2_\rho+\langle \overline{T} \rangle^2_\rho\right),
\eea
where $\langle \chi\rangle_\rho:= tr(\rho \chi)$ for $\chi=T,\bar T, \mathcal{A}$.  Note that $k_2$, $k_4$ and $k'_4$ are independent with $n$.

\subsection{Perturbative results of the functions $\mP$ and $\mP_l$}\label{shortsectionlargec}
If we only retain up to the second order, the result is similar with the vacuum case, requiring only the replacement of $b=\frac{c}{6}\log{\frac{l}{\epsilon}}$ with $b'=\frac{c}{6}\log{\frac{l}{\epsilon}}+k_2 l^2$ since their dependence on $n$ is the same
\bea
S^{(n)}=\frac{n+1}{n}(\frac{c}{6}\log{\frac{l}{a}}+k_2 l^2)+O(l^4).
\eea
At order $O(l^4)$ we will obtain more intriguing results. By definition we have
\bea
b:=-\log \lambda_m=\lim_{n\to \infty}S^{(n)}=\frac{c}{6}\log{\frac{l}{a}}+k_2l^2+k_4l^4+k'_4l^4,
\eea
where $\lambda_m$ is the maximal eigenvalue in this case.  Let us also define $b_0:=\frac{c}{6}\log{\frac{l}{a}}$.

By using (\ref{Inverse10}) and (\ref{Renyishort}), we obtain
\begin{align}\label{P4}
\mP(\lambda)=&\lambda^{-1}\Big[\frac{\sqrt{b_0}}{\sqrt{t}}I_1+\delta(t)+k_2I_0l^2+[(2k_4-10k'_4)I_0+\frac{1}{2}k_2^2 t^{\frac{1}{2}} b_0^{-\frac{1}{2}}I_1\nn\\
&+(-k_4+11k'_4)t b_0^{-1}I_2]l^4\Big],    
\end{align}
where the argument of $I_n$ is $2\sqrt{b_0t}$, and we have used 
\bea\label{calcu1}
\mL^{-1}\left[e^{\frac{b_0}{n}}n^k\right]=b_0^{(1+k)/2}t^{-(1+k)/2}I_{-(1+k)}(2\sqrt{b_0 t}).
\eea
Further, by using (\ref{Inverse1}) and (\ref{Renyishort}) we have
\begin{align}\label{thermalpl}
\mP_l(\lambda) 
=&-\frac{\p b}{\p l}\Big[\frac{c}{6}b_0^{-\frac{1}{2}}t^{\frac{1}{2}}I_1l^{-1}+\frac{\sqrt{b_0}}{\sqrt{t}}I_1+\delta(t)+[\frac{c}{6}k_2I_0+2k_2t^{\frac{1}{2}}b_0^{-\frac{1}{2}}I_1+\frac{c}{6}k_2tb_0^{-1}I_2]l\nn\\
&+[(-2k_2^2-\frac{c}{6}(2k_4-10k'_4))I_0+(-\frac{c}{6}\frac{1}{2}k_2^2+4(2k_4-10k'_4))t^{\frac{1}{2}}b_0^{-\frac{1}{2}}I_1\nn\\
&+(2k_2^2+\frac{c}{6}(3k_4-21k'_4))tb_0^{-1}I_2+(\frac{c}{6}\frac{1}{2}k_2^2-4(k_4-11k'_4))t^{\frac{3}{2}}b_0^{-\frac{3}{2}}I_3\nn\\
&-(\frac{c}{6}(k_4-11k'_4))t^{2}b_0^{-2}I_4]l^3\Big], 
\end{align}
where the argument of $I_n$ is $2\sqrt{b_0t}$.

It can be seen that the functions $\mP$ and $\mP_l$ depends on the expectation value $\langle \chi\rangle_\rho$. One could check they satisfy the relation
$\frac{\p \mP}{\p l}=-\frac{\p \mP_l}{\lambda}$.

One could also check the above results by the example of thermal state. Consider the thermal state with $\rho=e^{-\beta H}/Z(\beta)$. The R\'enyi entropy is given by 
\bea
S^{(n)}=\frac{c}{6}(1+\frac{1}{n})\log[\frac{\beta}{\pi\epsilon}\sinh(\frac{\pi l}{\beta})],
\eea
where $l$ represents the length of the interval, which is not necessarily assumed to be small.
With this result one could obtain $\mP(\lambda)$ and $\mP_l(\lambda)$ for the thermal state. The results are similar with the vacuum cases in section.\ref{examplevacuumsection}. Then one could expand the function $\mP(\lambda)$ and $\mP_l(\lambda)$ in term of $\frac{l}{\beta}$ in the region $l/\beta\ll 1$. The results should be same with (\ref{p4}) and (\ref{thermalpl}) up to $O(l^4)$ by using the expectation values of $\langle \chi\rangle_\beta:= tr (e^{-\beta H}\chi)/Z(\beta)$. The details of the calculations can be found in Appendix.\ref{shortsectionapp}.

\subsection{Zero point of $\mP_l$}
For the vacuum cases we find the function $\mP_L$ has one zero point which is given by $\lambda_m^2=e^{-2b_0}$. Here we would like to study the zero point of $\mP_l$ for the short interval case. We assume the zero point is given by the form $\tilde{\lambda}_0=e^{-\tilde{t}_0-b}$, with  $\tilde{t}_0=b_0+t_2l^2+t_4l^4+O(l^6)$, where $b_0$ is the zero point for the vacuum case. $t_2$ and $t_4$  should satisfy the following equations,   
\begin{align}
t_2&=k_2,\nn\\
t_4&=2k_4-10k'_4-\frac{48}{c}k_4\frac{I_2}{I_1}+\frac{528}{c}k'_4\frac{I_2}{I_1}-3k_4\frac{I_3}{I_1}+33k'_4\frac{I_3}{I_1},
\end{align}
where the argument of $I_n$ is $2b_0$.
For general theory the result is complicated. But for CFTs with holographic dual, we have  a large central charge $c$. For the function $I_n(x)$, since $\lim_{x\rightarrow\infty} I_n(x)=\frac{1}{\sqrt{2\pi x}}e^x$, in the limit $x\to \infty$ one would have $\lim_{x\to \infty} \frac{I_n(x)}{I_m(x)}=1$. Thus for large $c\gg 1$ , we have $b_0 \gg 1$ and 
\bea
\tilde{t}_0=b_0+k_2l^2-k_4l^4+23k'_4l^4+O(1/c).
\eea
By using (\ref{Renyishort})the EE is given by 
\bea 
S=\lim_{n\to 1}S^{(n)}=2b_0+2k_2l^2+24k'_4l^4+O(l^6).
\eea
It is remarkable that the zero point $\tilde{t}_0$ is associated with the EE $S$ and $b=S^{\infty}$,
\bea
\tilde{t}_0=S-b +O(1/c, l^6).
\eea
For the vacuum case we have $S=2b$, the zero point is given by $t_0=b$, which is consistent with the above results. One could also check the above relation for higher order of the short interval expansion. In the following section we will discuss the holographic CFTs. One would find the above relation is actually correct for arbitrary states that are dual to a bulk geometry. 
\section{Perturbation states}\label{section_perturbation}
Consider the density matrix
\bea
\rho=\rho_0+\delta \rho,
\eea
with the condition $tr \delta \rho=0$. One could obtain the R\'enyi entropy 
\begin{align}\label{perturbationSn}
S^{(n)}&=\frac{1}{1-n} \log(tr\rho^n) \nn\\
&=S^{(n)}(\rho_0)+\frac{n tr(\rho_0^{n-1}\delta \rho)}{(1-n)tr(\rho_0^n)}+O(\delta\rho^2)\nn\\
&=S^{(n)}(\rho_0)+\delta S^{(n)}+O(\delta\rho^2)
\end{align}
where we define
\bea\label{pertuabations1}
\delta S^{(n)}:=\frac{n tr(\rho_0^{n-1}\delta \rho)}{(1-n)tr(\rho_0^n)}.
\eea
In the following we will only keep the leading order of the perturbation.

On the other hand, by using $\rho_0=\sum_i\lambda_i^0 |\lambda_i^0\rangle \langle \lambda_i^0|$. 
We can rewrite (\ref{perturbationSn}) as
\bea\label{perturbationSn'}
\delta S^{(n)}:=\frac{n \sum_i {(\lambda_i^0)}^{n-1}\delta\lambda_i}{(1-n)\sum_i {(\lambda_i^0)}^{n}},
\eea
where we define
\begin{align}
\delta\lambda_i&:=\langle \lambda^0_i|\delta\rho|\lambda^0_i\rangle.
\end{align}

\subsection{Density of eigenstates}
Define $\lambda_m:= e^{-b}$ and $\lambda^0_m:=e^{-b_0}$, where $\lambda_m$ and $\lambda^0_m$ are maximal eigenvalues of $\rho$ and $\rho_0$ respectively. By definition we have $b=\lim_{n\to \infty}S^{(n)}(\rho)$, $b_0=\lim_{n\to \infty}S^{(n)}(\rho_0)$. Thus by using (\ref{perturbationSn'}), we have
\begin{align}\label{perturbation_b}
\lim_{n\to \infty}S^{(n)}(\rho)&=\lim_{n\to \infty}(S^{(n)}(\rho_0)+\delta S^{(n)})\nn\\
b&=b_0+\delta b,    
\end{align}
with
\begin{align}
\delta b&:=-\frac{\delta\lambda_m}{\lambda^0_m}\nn\\
\delta\lambda_m&:=\langle \lambda^0_m| \delta\rho |\lambda^0_m, \rangle,
\end{align}
where $|\lambda^0_m\rangle$ denotes the eigenstate for the maximal eigenvalue $\lambda^0_m$.
A useful form that we will utilize hereafter is $e^{\delta b}=1+\delta b=\frac{\lambda_m^0}{\lambda_m}$.

Now we are ready to evaluate the function $\mP$. By using the eq.(\ref{Inverse10}) we have
\bea\label{perturbationP}
&&\mP(e^{-b-t})= \lambda^{-1} \mL^{-1}\left[e^{nb+(1-n)S^{(n)}(\rho)}\right](t)\nn\\
&&\phantom{\mP(e^{-b-t})}=(\frac{\lambda^0_m}{\lambda_m}\lambda)^{-1}e^{\delta b} \mL^{-1}[e^{n b_0+(1-n)S^{(n)}(\rho_0)}](t)\nn\\
&&\phantom{\mP(e^{-b-t})=}+\lambda^{-1}\delta b\mathcal{L}^{-1}[n e^{n b_0+(1-n)S^{(n)}(\rho_0)}] (t)\nn\\
&&\phantom{\mP(e^{-b-t})=}+\lambda^{-1}\mL^{-1} [n e^{n b_0}tr(\rho_0^{n-1}\delta \rho)](t).
\eea
Due to the complexity of the above calculations, we will discuss the inverse Laplace transform in the above expression term by term.  Firstly, let us consider the first term.
Using $\lambda=e^{-b-t}$, since $e^{-b_0-t}=e^{-b-t}e^{\delta b}=\frac{\lambda^0_m}{\lambda_m}\lambda$, we have
\begin{align}\label{b1}
&(\frac{\lambda^0_m}{\lambda_m}\lambda)^{-1}e^{\delta b} \mL^{-1}[e^{n b_0+(1-n)S^{(n)}(\rho_0)}](t)\nn\\
&=(1+\delta b)(e^{-b_0-t})^{-1}\mL^{-1}[e^{n b_0+(1-n)S^{(n)}(\rho_0)}](t)\nn\\
&=\mP_0(\frac{\lambda^0_m}{\lambda_m}\lambda)+\delta b\mP_0(\frac{\lambda^0_m}{\lambda_m}\lambda),  
\end{align}

where $\mP_0$ denotes the density of eigenstate for the state $\rho_0$.
Note that in the last, we substituted the variable $\mP_0$ with $\frac{\lambda^0_m}{\lambda_m}\lambda$ to adjust for the range of values. The reason is easy to see in above mathematical calculations, since the inverse Laplace transform of $\mL^{-1}[e^{n b_0+(1-n)S^{(n)}(\rho_0)}](t)$ is $e^{-b_0-t}\mP_0(e^{-b_0-t})$. 

On the other hand, this adjustment can be also explained as follows: $\lambda$ in $\mP(\lambda)=\sum_i \delta(\lambda_i-\lambda)$ falls within the range $(0,\lambda_m]$, while $\lambda'$ in $\mP_0(\lambda')=\sum_i \delta(\lambda^0_i-\lambda')$ falls within the range $(0,\lambda^0_m]$. Hence, the purpose of this transformation is to ensure that both sides of the equation share the same variable range. 


Let us go on discussing the remaining two inverse Laplace transformation terms in (\ref{b1}).  By using the formula 
\bea\label{Laplaceproperty}
\mL^{-1}\{s \mL [f](s)\}(t)=\delta(t)f(0)+f'(t),
\eea
we have
\bea\label{perturbationP_2}
&&\lambda^{-1}\delta b\mathcal{L}^{-1}[n e^{n b_0+(1-n)S^{(n)}(\rho_0)}] (t)\nn\\
&&=\lambda^{-1}\delta b\mathcal{L}^{-1}[n \mL[e^{-b_0-t}\mP_0(e^{-b_0-t})]] (t)\nn\\
&&=\lambda^{-1}\delta b[\delta (t)e^{-b_0}\mP_0(e^{-b_0})+e^{-b_0-t}\frac{d\mP_0(e^{-b_0-t})}{d t}-e^{-b_0-t}\mP_0(e^{-b_0-t})]\nn\\
&&=\delta b\delta (t)\mP_0(e^{-b_0})-\delta b\lambda \mP'_0(\frac{\lambda^0_m}{\lambda_m}\lambda)-\delta b\mP_0(\frac{\lambda^0_m}{\lambda_m}\lambda),
\eea
where $\mP'_0(\lambda):=\frac{\p \mP_0(\lambda)}{\p \lambda}$.
It is worth noting that the final term in the expression above cancels out with the second term in formula (\ref{perturbationP}).

To calculate the last term of (\ref{perturbationP}), we define
\bea
\mP_{\delta}(\lambda'):=\sum_i \delta\lambda_i\delta(\lambda^0_i-\lambda'),
\eea
which can be taken as the average expectation value of the perturbation $\delta \rho$ in the eigenstates with eigenvalue $\lambda=e^{-b-t}$. It can be related to $tr(\rho_0^{n-1} \delta\rho)$ by 
\begin{align}
tr(\rho_0^{n-1} \delta\rho)&=\sum_i (\lambda^0_i)^{n-1} \delta\lambda_i\nn\\
&=\int_0^{\lambda^0_m}d\lambda' \lambda'^{n-1}\sum_i \delta\lambda_i\delta(\lambda^0_i-\lambda')\nn\\
&=\int_0^{\infty}dt e^{-n(b_0+t)} \mP_{\delta}(e^{-b_0-t}),     
\end{align}
where we use $\lambda':=e^{-b_0-t}$. One could obtain $\mP_{\delta}(e^{-b_0-t})$ by inverse Laplace transformation once $tr(\rho_0^{n-1}\delta \rho)$ is known. By using (\ref{pertuabations1}) and (\ref{perturbationSn'}), we have
\bea
\mP_{\delta}(e^{-b_0-t})&=\mL^{-1}\left[\frac{1-n}{n}e^{nb_0+(1-n)S^{(n)}(\rho_0)}\delta S^{n}\right],
\eea
the form of which is similar as $\mP_{\alpha_J}$ (\ref{Inverse1}).
Again use the formula (\ref{Laplaceproperty}), we obtain
\bea\label{perturbationP_3}
&&\lambda^{-1}\mL^{-1} [n e^{n b_0}tr(\rho_0^{n-1}\delta \rho)](t)\nn\\
&&=\lambda^{-1} [\delta (t)\mP_{\delta}(e^{-b_0})+\frac{d \mP_{\delta}(e^{-b_0-t})}{d t}] \nn\\
&&=\lambda^{-1}\delta (t)\mP_{\delta}(e^{-b_0})-\mP'_\delta(\frac{\lambda^0_m}{\lambda_m}\lambda).
\eea
where $\mP'_{\delta}(\lambda):=\frac{\p \mP_{\delta}}{\p  \lambda}$.
We compare the terms in (\ref{perturbationP_2}) and (\ref{perturbationP_3}) that contain $\delta(t)$ and find
\begin{align}\label{relationmaxperturbation}
\lambda^{-1}\delta (t)\mP_{\delta}(e^{-b_0})=-\delta b \delta (t) \mP_0(e^{-b_0}),
\end{align}
in the leading order of the perturbation. 

Combining all the aforementioned results, we derive the final expression. In summary, the function $\mP(\lambda)$ (\ref{perturbationP}) can be structured in the following form:
\bea\label{P_perturbation}
\mP(\lambda)=\mP(e^{-b-t})=\mP_0(e^{-b_0-t})+\delta \mP(e^{-b_0-t}),
\eea
with 
\begin{align}\label{twotermsperturbation}
&\mP_0(e^{-b_0-t})=\mP_0(\frac{\lambda^0_m}{\lambda_m}\lambda),\nn\\
&\delta\mP(e^{-b_0-t})=-\delta b\lambda\mP'_0(\frac{\lambda^0_m}{\lambda_m}\lambda)-\mP'_{\delta}(\frac{\lambda^0_m}{\lambda_m}\lambda).   
\end{align}

\subsection{Further discussion on the perturbation result}
\subsubsection{Normalization}
One could promptly verify that (\ref{P_perturbation}) complies with the normalization condition.  Through direct calculations we have
\bea\label{integralnorm}
&&\int_0^{\lambda_m} \lambda\mP_0(\frac{\lambda^0_m}{\lambda_m}\lambda)d\lambda
=1-2\delta b,\nn \\
&&-\int_0^{\lambda_m} \lambda\delta b\lambda\mP'_0(\frac{\lambda^0_m}{\lambda_m}\lambda)d\lambda
=-\delta b\frac{\lambda_m^3}{\lambda^0_m}\mP_0(\lambda^0_m)+2\delta b,\nn\\
&&-\int_0^{\lambda_m}\lambda\mP'_{\delta}(\frac{\lambda^0_m}{\lambda_m}\lambda) d\lambda=-\frac{\lambda^2_m}{\lambda^0_m}\mP_{\delta}(\lambda^0_m)+\frac{\lambda_m}{\lambda^0_m}\int_0^{\lambda^0_m}\mP_{\delta}(\lambda') d\lambda' =-\frac{\lambda^2_m}{\lambda^0_m}\mP_{\delta}(\lambda^0_m),\nn
\eea
where  in the last step we use the fact that $tr \delta\rho=\int_0^{\lambda^0_m}\mP_{\delta}(\lambda') d\lambda'=0$. 
Summing over the above results and using the relation (\ref{relationmaxperturbation}), we arrive at the normalization condition
\begin{align}
\int_0^{\lambda_m} \lambda\mP(\lambda)d\lambda=1.
\end{align}
\begin{figure}
\centering
\includegraphics[scale=0.4]{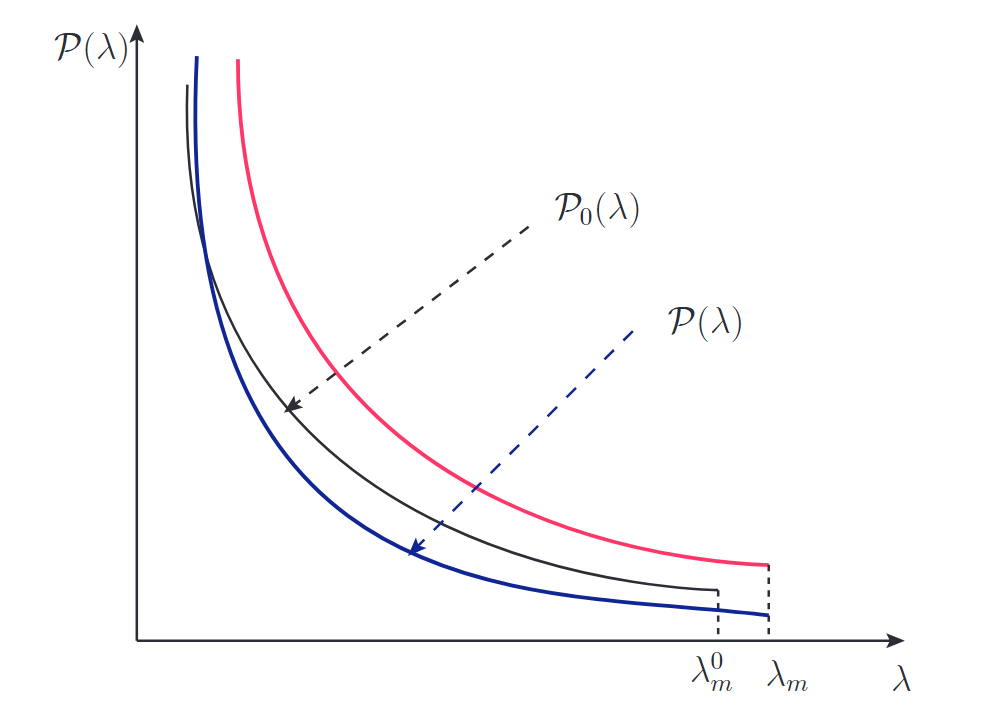}
\caption{Illustration of the function $\mP(\lambda)$ in the perturbation state. The black line is the unperturbed function $\mP_0$. The range of the eigenvalue would change under perturbation. The red line illustrates the perturbation of the function $\mP$ due to the alteration in the range of the variable $\lambda$. But it doesn't satisfy the normalization condition $\int_0^{\lambda_m} d\lambda \mP(\lambda)\lambda=1$. The blue one includes the  adjustment to satisfy the normalization. }
\label{plot2}
\end{figure}

Let's briefly analyze the implications of each term in the above expression (\ref{P_perturbation}). $\mP_0$ represents the unperturbed result of the density of eigenstates. As perturbation affect the the range of values of eigenvalues, i.e. the maximal eigenvalue changes from $\lambda^0_m$ to $\lambda_m$. The first term $\mP_0$  can be viewed as the change in the density distribution with the alteration in the range of the variable $\lambda$. As we can see from (\ref{integralnorm}) it doesn't satisfy the normalization. The second term in $\delta\mP(e^{-b_0-t})$ can be seen as the adjustment of the density distribution function itself to satisfy the normalization requirement. We illustrate the above explanation in the Fig.\ref{plot2}.

\subsubsection{The number of eigenvalues}
Since the function $\mP$ can be seen as the density of eigenstates, we can define the number of eigenstates larger than  $\lambda$ as
\begin{align}\label{numberdef}
n(\lambda)=\int_{\lambda}^{\lambda_m}\mP(\lambda')d\lambda'.
\end{align}

Taking (\ref{P_perturbation}) into the integration (\ref{numberdef}), we have
\begin{align}\label{number}
n(\lambda) =n_0(\frac{\lambda^0_m}{\lambda_m}\lambda)+\delta b\lambda\mP_0(\frac{\lambda^0_m}{\lambda_m}\lambda)+\mP_{\delta}(\frac{\lambda^0_m}{\lambda_m}\lambda),
\end{align}
where $n_0$ is the number of the unperturbed density matrix defined by
\begin{align}
n_0(\lambda)=\int_{\lambda}^{\lambda^0_m}\mP_0(\lambda')d\lambda'.
\end{align}
It's important to note that integration typically leads to divergence. For instance, in Section \ref{examplevacuumsection}, one can readily verify that the number of eigenstates is infinite. In essence, $N := n(\lambda=0)$ can be regarded as an approximation to the dimension of the density matrix $\rho_A$, which tends to be infinite in QFTs. Nevertheless, formally, we observe that the variation in dimension due to perturbations is linked to the function $\mP_\delta$, that is
\begin{align}
\Delta N := n(0)-n_0(0)=\mP_{\delta}(0).
\end{align}
Here the number of the maximal eigenvalue is well-defined and generally finite.  Taking $\lambda=\lambda$ into (\ref{number}) and using (\ref{relationmaxperturbation}) we obtain
\bea
n(\lambda_m)=n_0(\lambda^0_m),
\eea
which means the number of the maximal eigenvalue would be invariant at the leading order perturbation.

\subsection{The function $\mP_{\alpha_J}$}
It is straightforward to calculate the function $\mP_{\alpha}$ as we have done in previous section.
By using the formula (\ref{Inverse1}) we have
\bea
&&\mP_{\alpha_J}(e^{-b-t})=\mL^{-1}\left[\frac{1-n}{n}e^{nb+(1-n)S^{(n)}(\rho)}\frac{\p S^{(n)}(\rho)}{\p \alpha_J}\right]\nn \\
&&=\mL^{-1}\left[\frac{1-n}{n}e^{nb_0+(1-n)S^{(n)}(\rho_0)}(1+n\delta b+(1-n)\delta S^{(n)})\frac{\p (S^{(n)}(\rho_0)+\delta S^{(n)})}{\p \alpha_J}\right]\nn \\
&&=\mL^{-1}\left[\frac{1-n}{n}e^{nb_0+(1-n)S^{(n)}(\rho_0)}\frac{\p S^{(n)}(\rho_0)}{\p \alpha_J}\right]\nn \\
&&+\mL^{-1}\left[n\frac{1-n}{n}e^{nb_0+(1-n)S^{(n)}(\rho_0)}\frac{\p S^{(n)}(\rho_0)}{\p \alpha_J}\delta b\right]\nn \\
&&+\mL^{-1}\left[\frac{(1-n)^2}{n}e^{nb_0+(1-n)S^{(n)}(\rho_0)}\delta S^{(n)}\frac{\p S^{(n)}(\rho_0)}{\p \alpha_J}\right]\nn \\
&&+\mL^{-1}\left[\frac{1-n}{n}e^{nb_0+(1-n)S^{(n)}(\rho_0)}\frac{\p \delta S^{(n)}}{\p \alpha_J}\right]+\mL^{-1}\left[O(\delta\rho^2)\right].
\eea
Let us  discuss the four terms in the above results separately.

The first term is the function $\mP_{\alpha_J}$ for the unperturbed density matrix $\rho_0$, denote it by $\mP_{\alpha_J}^0(e^{-b_0-t})$. 
By using the formula (\ref{Laplaceproperty}), the second term is given by
\bea
&&\mL^{-1}\left[n\frac{1-n}{n}e^{nb_0+(1-n)S^{(n)}(\rho_0)}\frac{\p S^{(n)}(\rho_0)}{\p \alpha_J}\right] \delta b \nn \\
&&=\mL^{-1}\left[n \mL[\mP_{\alpha_J}^0(e^{-b_0-t})]\right] \delta b\nn \\
&&=  \delta b ( \delta(t) \mP_{\alpha_J}^0(e^{-b_0})+\frac{d\mP_{\alpha_J}^0(e^{-b_0-t})}{dt} )\nn\\
&&=  \delta b \delta(t) \mP_{\alpha_J}^0(e^{-b_0})-\delta b \lambda \mP_{\alpha_J}^{0'}\left(\frac{\lambda^0_m}{\lambda_m}\lambda\right),
\eea
where $\mP_{\alpha_J}^{0'}:=\frac{\p\mP_{\alpha_J}^0}{\p\lambda}$. The third term can be simplified as follows:
\bea\label{thirdterm}
&&\mL^{-1}\left[\frac{(1-n)^2}{n}e^{nb_0+(1-n)S^{(n)}(\rho_0)}\delta S^{(n)}(\rho_0)\frac{\p S^{(n)}(\rho_0)}{\p \alpha_J}\right]\nn\\
&&=\mL^{-1}\left[\frac{(1-n)^2}{n}e^{nb_0+(1-n)S^{(n)}(\rho_0)}\frac{n tr(\rho_0^{n-1}\delta \rho)}{(1-n)tr(\rho_0^n)}\frac{\p S^{(n)}(\rho_0)}{\p \alpha_J}\right]\nn\\
&&=\mL^{-1}\left[(1-n)e^{nb_0}tr(\rho_0^{n-1}\delta \rho)\frac{\p S^{(n)}(\rho_0)}{\p \alpha_J}\right].
\eea
One could further simplify the above term if the R\'enyi entropy $S^{(n)}(\rho_0)$ is given. Similar terms will also appear in the fourth term, albeit with opposite signs, allowing them to cancel each other out. The fourth term is considerably more intricate. In order to articulate the outcomes, we need to introduce the following quantities:
\begin{align}
\mP_{(\delta \alpha_J)}(\lambda')&:= \sum_i \frac{\p \delta \lambda_i}{\p \alpha_J}\delta(\lambda^0_i-\lambda')\\
\mP_{(\delta)(\alpha_J)}(\lambda')&:=\sum_i \delta \lambda_i\frac{\p \lambda^0_i}{\p \alpha_J}\delta(\lambda^0_i-\lambda').    
\end{align}
These two functions illustrate the relationship between the variation $\delta\lambda_i$ and the parameter $\alpha_J$. They can be computed once $\frac{\partial \text{tr}(\rho_0^{n-1} \delta\rho)}{\partial \alpha_J}$ is known. With these definitions we have
\begin{align}
&\mL^{-1}\left[\frac{1-n}{n}e^{nb_0+(1-n)S^{(n)}(\rho_0)}\frac{\p \delta S^{(n)}(\rho_0)}{\p \alpha_J}\right]\nn \\
&=\mL^{-1}\left[\frac{1-n}{n}e^{nb_0+(1-n)S^{(n)}(\rho_0)}\frac{\p }{\p \alpha_J} \frac{n tr(\rho_0^{n-1}\delta \rho)}{(1-n)tr(\rho_0^n)}\right]\nn \\
&=\mL^{-1}\left[e^{nb_0}\frac{\p }{\p \alpha_J}[tr(\rho_0^{n-1}\delta \rho)]\right]
-\mL^{-1}\left[e^{nb_0}tr(\rho_0^{n-1}\delta \rho)(1-n)\frac{\p S^{(n)}(\rho_0)}{\p \alpha_J}\right]\nn \\
&=\mL^{-1}\left[e^{nb_0}\frac{\p }{\p \alpha_J}[\sum_i (\lambda^0_i)^{n-1}\langle i|\delta\rho|i\rangle]\right]-\mL^{-1}\left[(1-n)e^{nb_0}tr(\rho_0^{n-1}\delta \rho)\frac{\p S^{(n)}(\rho_0)}{\p \alpha_J}\right]\nn \\
&=\mL^{-1}\left[e^{nb_0} \sum_i (\lambda^0_i)^{n-1}\frac{\p \langle i | \delta \rho| i\rangle}{\p \alpha_J} \right]+\mL^{-1}\left[e^{nb_0}\sum_i (n-1)(\lambda^0_i)^{n-2}\frac{\p \lambda^0_i}{\p \alpha_J}\langle i|\delta\rho|i\rangle\right] \nn \\
&-\mL^{-1}\left[(1-n)e^{nb_0}tr(\rho_0^{n-1}\delta \rho)\frac{\p S^{(n)}(\rho_0)}{\p \alpha_J}\right]\nn \\
&=\mL^{-1}\left[ \mL[\mP_{(\delta \alpha_J)}(e^{-b_0-t})] \right]
+\mL^{-1}\left[ (n-1)\mL[\mP_{(\delta)(\alpha_J)}(e^{-b_0-t})e^{b_0+t}] \right] \nn \\
&-\mL^{-1}\left[(1-n)e^{nb_0}tr(\rho_0^{n-1}\delta \rho)\frac{\p S^{(n)}(\rho_0)}{\p \alpha_J}\right]\nn \\
&=\mP_{(\delta \alpha_J)}(e^{-b_0-t})+\delta(t)e^{b_0}\mP_{(\delta)(\alpha_J)}(e^{-b_0})+[e^{b_0+t}\frac{d}{dt}\mP_{(\delta)(\alpha_J)}(e^{-b_0-t})+e^{b_0+t}\mP_{(\delta)(\alpha_J)}(e^{-b_0-t})]\nn \\
&-e^{b_0+t}\mP_{(\delta)(\alpha_J)}(e^{-b_0-t})-\mL^{-1}\left[(1-n)e^{nb_0}tr(\rho_0^{n-1}\delta \rho)\frac{\p S^{(n)}(\rho_0)}{\p \alpha_J}\right]\nn\\
&=\mP_{(\delta \alpha_J)}(\frac{\lambda^0_m}{\lambda_m}\lambda)+\delta(t)e^{b_0}\mP_{(\delta)(\alpha_J)}(e^{-b_0})-\mP'_{(\delta)(\alpha_J)}(\frac{\lambda^0_m}{\lambda_m}\lambda)\nn \\
&-\mL^{-1}\left[(1-n)e^{nb_0}tr(\rho_0^{n-1}\delta \rho)\frac{\p S^{(n)}(\rho_0)}{\p \alpha_J}\right], 
\end{align}
where in the derivation we use the formula (\ref{Laplaceproperty}) again, and define $\mP'_{(\delta)(\alpha_J)}(\lambda):=\frac{\p\mP_{(\delta)(\alpha_J)}}{\p\lambda}$.  As mentioned earlier, the final term in the above results exactly cancels out with the result from the third term (\ref{thirdterm}).

We also notice that
\begin{align}
\delta(t)e^{b_0}\mP_{(\delta)(\alpha_J)}(e^{-b_0})
=-\delta(t)\delta b \mP^0_{\alpha_J}(e^{-b_0}).
\end{align} 
In a summary, the final result can be expressed in the following concise form
\bea\label{PaJ_perturbation}
\mP_{\alpha_J}(e^{-b-t})=\mP^0_{\alpha_J}(e^{-b_0-t})+\delta \mP_{\alpha_J}(e^{-b_0-t})
\eea
with
\bea
\mP^0_{\alpha_J}(e^{-b_0-t})=\mP^0_{\alpha_J}(\frac{\lambda^0_m}{\lambda_m}\lambda)
\eea
and
\begin{align}
\delta \mP_{\alpha_J}(e^{-b_0-t})=-\delta b\lambda \mP_{\alpha_J}^{0'}(\frac{\lambda^0_m}{\lambda_m}\lambda)+\mP_{(\delta \alpha_J)}(\frac{\lambda^0_m}{\lambda_m}\lambda)-\mP'_{(\delta)(\alpha_J)}\left(\frac{\lambda^0_m}{\lambda_m}\lambda\right).    
\end{align}
The aforementioned result for $\mP_{\alpha_J}(\lambda)$ is very similar to $\mP(\lambda)$. Similar explanations can be provided for each term in (\ref{PaJ_perturbation}), akin to what we have done in the previous section. 

We can also check that $\mP$ and $\mP_{\alpha_J}$ satisfy the consistent relation (\ref{consistent1}). See the Appendix.\ref{Consistentcheck_perturbation} for the details.

\subsection{A simple example of perturbation states}
Let's consider the example studied in section.\ref{examplevacuumsection}. Suppose the density matrix $\rho_0$ corresponds to the interval with length $l_0$. While the density matrix $\rho=\rho_0+\delta \rho$ corresponds to the interval $l=l_0+\delta l$ with $\delta \ll l$. Thus we can take $\delta \rho$ as perturbation and obtain
\begin{align}
&\delta S^{(n)}=\frac{c}{6}(1+\frac{1}{n})\frac{\delta l}{l_0}\\
&\delta b=b-b_0=\frac{c}{6}\frac{\delta l}{l_0}
\end{align}
Now it is straightforward to obtain the function $\mP_\delta$ 
\begin{align}
\mP_{\delta}(e^{-b_0-t})&=\mL^{-1}[\frac{(1-n)}{n}e^{nb_0}e^{(1-n)S^{(n)}(\rho_0)}\delta S^{(n)}](t)\nn\\
&=-\frac{c}{6}\frac{\delta l}{l_0}\left[\frac{(b_0-t) I_1\left(2 \sqrt{b_0t}\right)}{\sqrt{b_0 t}}+\delta (t)\right].
\end{align}
Further using  (\ref{P_perturbation}) we have 
\begin{align}\label{perturexam}
\mP(\lambda)&=\mP_0(\frac{\lambda^0_m}{\lambda_m}\lambda)-\delta b\lambda \mP'_0(\frac{\lambda^0_m}{\lambda_m}\lambda)-\mP'_{\delta}(\frac{\lambda^0_m}{\lambda_m}\lambda)\nn\\
&=\frac{1}{\lambda}\frac{\sqrt{b_0} I_1}{\sqrt{-b_0-\log \lambda'}}+\delta (\lambda_m-\lambda)+\frac{c}{6}\frac{\delta l}{l_0}[\frac{I_1}{\lambda'\sqrt{(-b_0-\log \lambda')b_0}}+\frac{I_2}{\lambda'}],
\end{align}
where $\lambda':=\frac{\lambda^0_m}{\lambda_m}\lambda$.
On the other hand, we can also expand $\mP(\lambda)$ in (\ref{spectra1}) and keep the first order $O(\frac{\delta l}{l_0})$. The result is same as (\ref{perturexam}). This can be seen as a consistent check of our general result (\ref{P_perturbation}).

In this example we can also check the number of eigenstates (\ref{number}). With some calculations we have
\begin{align}
n(e^{-b-t})&=n_0(e^{-b_0-t})+\delta b \lambda \mP_0(e^{-b_0-t})+\mP_{\delta}(e^{-b_0-t})\nn\\
&=n_0(e^{-b_0-t})+\frac{\sqrt{t}I_1(2\sqrt{b_0t})}{\sqrt{b_0}}\delta b,
\end{align}
where $n_0(e^{-b_0-t})=I_0(2\sqrt{b_0t})$. Let us consider the two limit $\lambda\to 0$ and $\lambda\to \lambda_m$
We care about $n(0)$ and $n(\lambda_m)$. As expected $n(0)$ is divergent. Formally, we have
\bea
N=n(0)-n_0(0)=\mP_{\delta}(0)=\lim_{t\to\infty}\frac{\sqrt{t}I_1(2\sqrt{b_0t})}{\sqrt{b_0}}\delta b.
\eea
On the other hand, we also find
\begin{align}
n(\lambda_m)=n_0(\lambda^0_m)=I_0(0)=1,
\end{align}
which is consistent with our previous discussion, the number of maximal eigenvalues does not change at the first order perturbation. We anticipate that this outcome holds true beyond the leading order perturbation, owing to the presence of a Dirac delta function $\delta(\lambda-\lambda_m)$ in the density of eigenstates $\mP(\lambda)$. In this example, the maximal eigenstate corresponds to the vacuum state on the annulus following the conformal transformation (\ref{conformalmap}). It's natural for the vacuum state to be non-degenerate in this context.Our results indicate that this non-degeneracy exhibits robustness under first-order perturbations.

\section{Geometric states in holographic theory}\label{section_holographic}
In this section, our focus will be on holographic theory, a framework wherein certain special states can be effectively described by classical geometry in the semi-classical limit $G\to 0$. These states are referred to as geometric states. In the semi-classical limit $G\to 0$, quantum fluctuations are suppressed. Certain non-local observables, such as entanglement entropy, may have a bulk geometric dual. Entanglement can be utilized as a probe to determine whether a given state can be dual to bulk geometry. In fact, constructing states that cannot be dual to bulk geometry is not a difficult task; refer to \cite{Guo:2018fnv} for details. 
\subsection{The functions $\mP$ and $\mP_{\alpha_J}$}
The density of eigenstates $\mP$ also has some interesting feature. By the formula (\ref{Inverse10}) one have
\bea
\mP(e^{-b-t})e^{-b-t}=\mL^{-1}\left[e^{n b+(1-n)S^{(n)}}\right]=\frac{1}{2\pi i} \int_{\gamma_0-i\infty}^{\gamma_0+i\infty}dn e^{s_n},
\eea
where
\bea
s_n:=n(t+b)+(1-n)S^{(n)}.
\eea
One important feature of geometric state is that the R\'enyi entropy has a gravity dual, follows the area law formula\cite{Dong:2016fnf}, 
\bea
n^2\p_n\left(\frac{n-1}{n}S^{(n)} \right)=\frac{\mathcal{B}_n}{4G},
\eea
where $\mathcal{B}_n$ denotes the area of the bulk codimension-2 brane. The tension of the brane is related to the index $n$ by $T_n=\frac{n-1}{4nG}$. In the limit $n\to 1$ we would obtain the Ryu-Takayanagi formula. 

Note that by the above holographic formula, one can see the R\'enyi entropy for geometric state should be of $O(1/G)$. This permits us to use saddle point approximation to evaluate the functions $\mP$ and $\mP_{\alpha_J}$. 
By using saddle point approximation and the holographic R\'enyi entropy formula, one could derive 
\begin{align}
\mP(e^{-b-t})e^{-b-t}=\frac{1}{2\pi}\sqrt{\frac{2\pi}{\frac{\p^2 s_n}{\p n^2}|_{n=n*}}}e^{\frac{\mathcal{B}_{n^*}}{4G}}.
\end{align}
where $\mathcal{B}_{n^*}$ is the area of the cosmic brane with tension $\mu=\frac{n^*-1}{4G n^*}$, $n^*$ is determined by the saddle point condition
\bea
\p_n s_n|_{n=n^*}=0.
\eea
We can proceed with the evaluation of the function $\mP_{\alpha_J}$ for the geometric state. It needs to evaluate the inverse Laplace transformation (\ref{Inverse1}), that is
\bea
\mP_{\alpha_J}(e^{-b-t})=\frac{1}{2\pi i}\int_{\gamma_0-i\infty}^{\gamma_0 +i\infty}dn e^{s_{\alpha_J,n}}
\eea
where 
\bea
s_{\alpha_J,n}:=nt+nb+(1-n)S^{(n)}+\log \frac{\p S^{(n)}}{\p \alpha_J}+\log \frac{1-n}{n}.
\eea
For geometric states we expect the R\'enyi entropy should be of order $ O(1/G)$.  Thus in the semi-classical limit the last two logarithmic term can be ignored. Therefore, the saddle point approximation equation 
\bea
\p_n s_{\alpha_J,n}\simeq \p_n s_{n}=0,
\eea
holds at the leading order of $O(1/G)$. The solution is given by $n^*$. Taking $n^*$ back into $s_{\alpha_J,n}$, the function $\mP_{\alpha_J}(e^{-b-t})$ can be approximated by
\bea
\mP_{\alpha_J}(e^{-b-t})\simeq \frac{1}{2\pi}\sqrt{\frac{2\pi}{\frac{\p^2 s_{\alpha_J,n}}{\p n^2}|_{n=n*}}} e^{s_{\alpha_J,n^*}}\propto e^{\frac{\mathcal{B}_{n^*}}{4G}}\frac{\p S^{(n^*)}}{\p \alpha_J}\frac{1-n^*}{n^*}.
\eea
Let's analyze the zero point of the function $\mP_{\alpha_J}$.
The function $\frac{\partial S^{(n)}}{\partial \alpha_J}$ generally does not vanish except at certain special values of $\alpha_J$. For instance, considering $\alpha_J$ as the size of the subsystem denoted by $L$. It can be demonstrated that for a pure state, $\frac{\partial S^{(n)}}{\partial L}$ would vanish if $L$ corresponds to half of the entire system size, refer to the discussions in the next section. Here, we assume that it is not at this critical point. Therefore, the zero point of the function $\mP_{\alpha_J}$ is given by $n^*=1$. The equation 
\bea
\p_n s_n|_{n=n^*=1}=0,
\eea
gives the solution $t=S-b$ or equally $t=S-S^{\infty}$, where $S$ is entanglement entropy. 

In Section \ref{shortsectionlargec}, we evaluate the function $\mP_{L}$ and discover that its zero point is also determined by $t=S-b$ in the large $c$ limit. In fact, for a holographic theory, the central charge $c \sim O(1/G)$. Hence, the large $c$ limit precisely corresponds to the semi-classical limit. The zero point of $\mP_L$ in the large $c$ limit discussed in Section \ref{shortsectionlargec} serves as a non-trivial validation of our general conclusion. It's important to note that our findings in this section hold true for arbitrary parameters $\alpha_J$, except for the parameters $c$ or $G$. In the paper \cite{Guo:2020roc}\cite{Guo:2021tzs}, the authors construct the so-called fixed area state in CFTs. It's also noteworthy that the fixed area state with $t=S-b$ is indeed quite special, as it can be considered as an approximate state for the reduced density matrix $\rho_A$ in the large $c$ limit. There might exist profound connections between these intriguing results.
\subsection{Higher dimension example}
In \cite{Hung:2011nu}, the authors consider the holographic R\'enyi entropy for a sphere region in $d$-dimensional spacetime. For the dual bulk theory being Einstein gravity,  the holographic R\'enyi Entropy is
\begin{align}
S_d^{(n)}=\frac{n}{n-1}\pi V_{\Sigma}(\frac{\tilde{L}}{l_P})^{d-1}(2-x_n^{d-2}(1+x_n^2)),
\end{align}
where $x_n=\frac{1}{dn}(1+\sqrt{1-2dn^2+d^2n^2})$, $V_{\Sigma}$ denote the ‘coordinate’ volume of the hyperbolic plane, $l_P^{d-1}=8\pi G$, and $\tilde{L}^2$ gives the AdS curvature scale. Thus we have
\begin{align}\label{holographic_Sb}
S&=\lim_{n\to1}S_d^{(n)}=2\pi V_{\Sigma}(\frac{\tilde{L}}{l_P})^{d-1},\nn\\
b&=\lim_{n\to\infty}S_d^{(n)}=2\pi V_{\Sigma}(\frac{\tilde{L}}{l_P})^{d-1}(1-\frac{d-1}{d}(\frac{d-2}{d})^\frac{d-2}{2}).
\end{align}
 $S_d^{(n)}$ can be rewritten as
\begin{align}
S_d^{(n)}=\frac{n}{n-1}(1-\frac{1}{2}x_n^{d-2}(1+x_n^2))S.
\end{align}
We care about the zero point of the function $\mP_{\alpha_J}$. In principle one could obtain the expression of $\mP_{\alpha_J}$ by directly computing the inverse Laplace transformation (\ref{Inverse1}). However, we cannot obtain an analytical result. Since we are considering the holographic theory, one could use the saddle point approximation. It can be found that the zero point is given by $t=S-b$.
And further, by using (\ref{holographic_Sb}), the zero point can be written as
\begin{align}
 t=\left[(1-\frac{d-1}{d}(\frac{d-2}{d})^\frac{d-2}{2})^{-1}-1\right]b.
\end{align}

\begin{figure}
\centering
\includegraphics[scale=0.6]{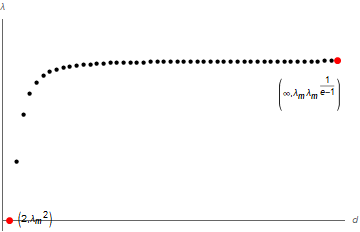}
\caption{Illustration of the zero point $\lambda$ varies with dimension $d$.}
\label{zero}
\end{figure}

The corresponding zero point $\lambda_0$ is given by
\begin{align}
\lambda_0=\lambda_m^{(1-\frac{d-1}{d}(\frac{d-2}{d})^\frac{d-2}{2})^{-1}},
\end{align}
where $\lambda_m=e^{-b}$. If $d=2$, $\lambda_0=\lambda_m^2$, which is consistent with the result in section.\ref{examplevacuumsection}. In the limit $d\to\infty$, the zero point $\lambda_0\to \lambda_m\lambda_m^{\frac{1}{e-1}}$. It can be shown $\lambda_0$  is a monotonically increasing function of $d$. We plot the zero point $\lambda_0$ as a function of the dimension $d$ in Fig.\ref{zero}.

\section{Entanglement entropy and the functions}\label{section_7}
In the preceding sections, we explored the properties of functions like $\mP$, $\mP_{\alpha_J}$, and others. These functions inherently encompass more information than just the entanglement measure, such as the entanglement entropy. In this section, our aim is to demonstrate how the properties of these functions directly correlate with certain aspects of entanglement entropy.
\subsection{First derivative of entanglement entropy with respect to the subsystem size}
Recall the definition of entanglement entropy 
\bea
S=-\sum_i \lambda_i \log \lambda_i.
\eea
The dependence of entanglement entropy on certain parameters is directly related to those functions we have previously studied. Let us focus on the size of the subsystem $L$. We have
\bea\label{firstderivative}
&&\frac{\p S}{\p L}=-\sum_i \frac{\p \lambda_i}{\p L}\log\lambda_i-\sum_i \frac{\p \lambda_i }{\p L},\nn \\
&&\phantom{\frac{\p S}{\p L}}=-\int_{0}^{\lambda_m}d\lambda \mP_L(\lambda) \log\lambda,
\eea
where we have used the fact $\sum_i \frac{\p \lambda_i }{\p L}=0$.
\begin{figure}
\centering
\includegraphics[scale=0.47]{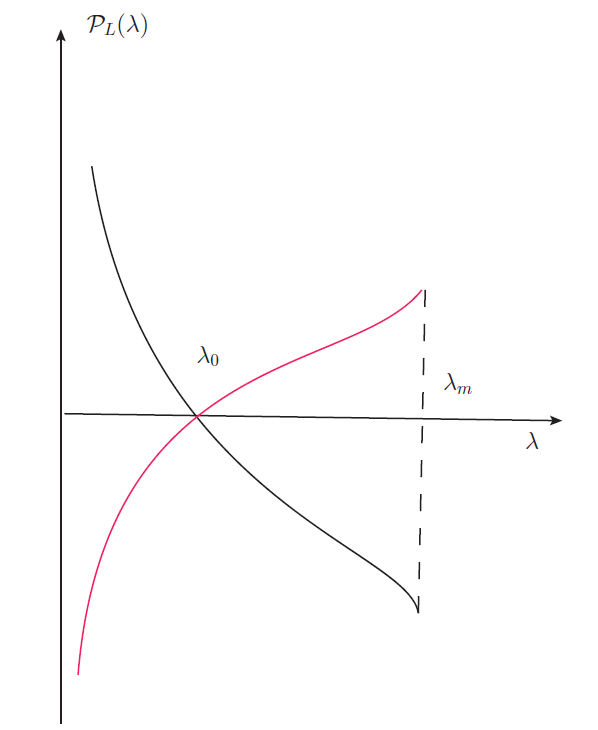}
\caption{Illustration of the function $\mP_L$. The black line and red line are two typical functions for $\mP_L$. $\lambda_0$ is the zero point of $\mP_L$, $\lambda_m$ is the maximal eigenvalue. }
\label{plot1}
\end{figure}
Support the function $\mP_L$ is given by the black line shown in Fig.\ref{plot1}, which is similar to the example of a single interval in the vacuum state on an infinite line (see Fig.\ref{plot_P}).  There is a zero point $\lambda_0$, $\mP_L>0$ for $\lambda<\lambda_0$ and $\mP_L<0$ for $\lambda_0<\lambda \le \lambda_m$. It can be shown that 
\bea
&&\frac{\p S}{\p L}=-\int_{0}^{\lambda_0}d\lambda \mP_L(\lambda) \log\lambda-\int_{\lambda_0}^{\lambda_m}d\lambda \mP_L(\lambda) \log\lambda\nn \\
&&\phantom{\frac{\p S}{\p L}}\ge -\int_{0}^{\lambda_0}d\lambda \mP_L(\lambda) \log\lambda_0-\int_{\lambda_0}^{\lambda_m}d\lambda \mP_L(\lambda) \log\lambda\nn \\
&&\phantom{\frac{\p S}{\p L}}=\int_{\lambda_0}^{\lambda_m}d\lambda \mP_L(\lambda) \log\lambda_0-\int_{\lambda_0}^{\lambda_m}d\lambda \mP_L(\lambda) \log\lambda\nn \\
&&\phantom{\frac{\p S}{\p L}}=\int_{\lambda_0}^{\lambda_m}d\lambda \mP_L(\lambda) \log\frac{\lambda_0}{\lambda}\ge 0,
\eea
where in the first step we use $-\log\lambda \ge -\log \lambda_0$ for $0<\lambda<\lambda_0$, in the second step we use $\sum_i \frac{\p \lambda_i }{\p L}=\int_0^{\lambda_m}d\lambda \mP_L(\lambda)=0$, in the last step $\mP_L\le 0$ and $\log\frac{\lambda_0}{\lambda}\le 0$ for $\lambda_0\le \lambda\le \lambda_m$. While if the function $\mP_L$ is like the form of the red line in Fig.\ref{plot1}. We can demonstrate, as previously done, that $\frac{\partial S}{\partial L}\le 0$.

The above discussion show that whether the entanglement entropy $S$ increases or decreases with the increase of L depends on the characteristics of function $\mP_L$. In section.\ref{sectioncylinder} we obtain the function $\mP_L$ for one interval with length $L$ on cylinder with  circumference $R$. $\mP_L$ is taken as the form of the black line in Fig.\ref{plot1} for $L<\frac{R}{2}$. This shows that $S$ is monotonically increasing function
of $L$ in this region. While  $S$ is monotonically decreasing function
of $L$ in the region $\frac{R}{2}<L<R$. There is a critical point $L=\frac{R}{2}$, where $\frac{\p 
 S}{\p L}=0$. At this point we also have $\mP_L=0$, since $\cot{\frac{\pi}{2}}=0$.

 For one interval in arbitrary pure state, say $|\psi\rangle$, on cylinder, we have $S(R-L)=S(L)$, which leads to 
 \bea\label{purerelation}
-S'(R-L)=S'(L).
 \eea
 Thus one could obtain $\frac{\p S}{\p L}|_{L=\frac{R}{2}}=0$. At the point $L=\frac{R}{2}$ we expect the function $\mP_{L}=0$. By utilizing (\ref{purerelation}), one can observe that the sign of $\frac{\partial S}{\partial L}$ differs between the two cases: $L<\frac{R}{2}$ and $L>\frac{R}{2}$. Hence, we anticipate that the function $\mP_L$ would resemble the black and red lines depicted in Fig. \ref{plot1} for $L<\frac{R}{2}$ and $L>\frac{R}{2}$, respectively. 
The above assertions can be verified through specific explicit examples.

\begin{figure}
\centering
\subfigure[]{\includegraphics[scale=0.48]{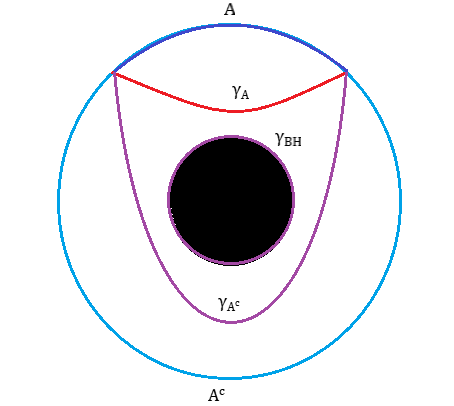}}
\subfigure[]{\includegraphics[scale=0.48]{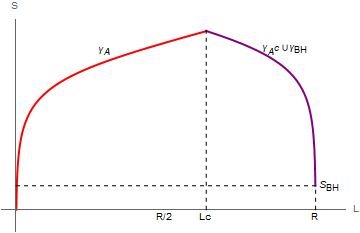}}
\caption{(a) shows two types of minimal surfaces $\gamma_A$ and $\gamma_{A^c}\cup\gamma_{BH}$, which are shown in red and purple line, respectively. We notice that $\gamma_A$ is homotopy with $\gamma_{A^c}\cup\gamma_{BH}$ instead of $\gamma_{A^c}$ in this case. Since the result should take the minimum value in the extreme surfaces, for $L<L_{\text{c}}$ the $S$ is given by $\gamma_A$, whereas for $L>L_{\text{c}}$ the $S$ is given by $\gamma_{A^c}\cup\gamma_{BH}$, which are shown in red and purple respectively in (b).}
\label{Plot_BTZ}
\end{figure}

One more interesting example is one interval in thermal state with $\beta$ on cylinder with circumference $R$. For high temperature limit $R/\beta$ the gravity dual is described by BTZ black hole. By using the RT formula, one could directly evaluate the holographic entanglement entropy by choosing the global minimal surface. It has been demonstrated that the holographic entanglement entropy undergoes a phase transition at a critical point $L=L_{\text{c}}$. These two phases correspond to distinct types of minimal surfaces, illustrated in Fig.\ref{Plot_BTZ}. For $L<L_{\text{c}}$, it is observed that $S$ is a monotonically increasing function of $L$, whereas for $L>L_{\text{c}}$, it behaves as a monotonically decreasing function. Based on our earlier discussions, we can conclude that at the critical point $L_{\text{c}}$, $\mP_{L_{\text{c}}}=0$. Consequently, the function $\mP_L$ serves as a means to identify the phase transition of entanglement entropy.

\subsection{Second derivative of entanglement entropy with respect to the subsystem size}
 In \cite{Casini:2004bw} the author introduce the so-called entropy c-function $c(L):=L\frac{\p S}{\p L}$. By the combination of the Lorentz symmetry and the strong subadditivity of entropy, it can be shown that $c'(L)\le 0$.  This prompts us to consider the quantity $c'(L)=L\frac{\p }{\p L} \left(L\frac{\p  S}{\p L}\right)$ and its relation to the functions $\mP_{L}$, $\mP_{L^2}$ and $\mP_{L_2}$. 
By definition we have
\bea\label{secondderivative}
&&c'(L)=-L\sum_i \frac{\p \lambda_i}{\p L} \log \lambda_i -L^2\left(\sum_i \frac{\p^2 \lambda_i}{\p L^2}\log\lambda_i+\sum_i \left(\frac{\p\lambda_i}{\p L}\right)^2\lambda_i^{-1} \right) \nn \\
&&\phantom{c'(L)}=-L \int_{0}^{\lambda_m} d\lambda \mP_L \log \lambda-L^2\left(\int_0^{\lambda_m}d\lambda\mP_{L_2}\log \lambda+\int_{0}^{\lambda_m}d\lambda \mP_{L^2}\lambda^{-1} \right).\nn
\eea
Firstly, let us consider the contribution from the maximal eigenvalue $\lambda_m$. Suppose the maximal eigenvalue is non-degenerate.  Its contribution is given by
\bea
&&-L \frac{\p \lambda_m}{\p L} \log \lambda_m -L^2\left( \frac{\p^2 \lambda_m}{\p L^2}\log\lambda_m+ \left(\frac{\p\lambda_m}{\p L}\right)^2\lambda_m^{-1} \right)\nn \\
&&=-L\left(1+L\frac{\p }{\p L}\right)\left(\frac{\p \lambda_m}{\p L}\log \lambda_m \right).
\eea
These terms come from the Dirac delta $\delta(\lambda-\lambda_m)$ in the functions $\mP_L$, $\mP_{L^2}$ and $\mP_{L_2}$. Let us define the functions without the Dirac delta terms by $\tilde{\mP}_L$, $\tilde{\mP}_{L^2}$ and $\tilde{\mP}_{L_2}$. Generally, the functions can be written as
\bea
&&\mP_L =\tilde{\mP}_L+\frac{\p \lambda_m}{\p L} \delta(\lambda_m-\lambda),\nn \\
&&\mP_{L^2} =\tilde{\mP}_{L^2}+\left( \frac{\p \lambda_m}{\p L}\right)^2 \delta(\lambda_m-\lambda),\nn \\
&&\mP_{L_2}=\tilde{\mP}_{L_2}+\frac{\p^2 \lambda_m}{\p L^2} \delta(\lambda_m-\lambda).
\eea
By using (\ref{r2}) with $\alpha_J=L$, the contribution from other eigenvalues is given by
\bea
-L \int_0^{\lambda_m} d\lambda \tilde{\mP}_L \log\lambda-L^2 \int_0^{\lambda_m}d\lambda \frac{\p \tilde{\mP}_L}{\p L}\log\lambda-L^2 \tilde{\mP}_{L^2} \log\lambda |^{\lambda_m}_0,
\eea
where the last term is the boundary term at the eigenvalues $\lambda_m$ and  $0$. In summary, $c'(L)$ can be expressed as
\bea\label{secondfinal}
c'(L)=-L\int_0^{\lambda_m}d\lambda(1+L\frac{\p }{\p L}) \mP_{L} \log\lambda-L^2 \tilde{\mP}_{L^2} \log\lambda |^{\lambda_m}_0.
\eea
Let us consider the boundary term $\lim_{\lambda\to 0}L^2 \tilde{\mP}_{L^2} \log\lambda$, which can be determined by studying the behavior of $\mP_{L^2}$ as $\lambda$ approaches zero. However, our knowledge about the properties of this function is limited. An explicit example might be found in the case of a single interval in the vacuum state (see (\ref{examplevacuumsection})). Using equation (\ref{diffferential}), we find that $\mP_{L^2}(\lambda)\sim \lambda(\log\lambda)^2 e^{2\sqrt{-b\log \lambda}}$ as $\lambda$ approaches zero. For this particular example, we observe that $\lim_{\lambda\to 0}L^2 \tilde{\mP}_{L^2} \log\lambda=0$. Starting from  (\ref{secondderivative}), $c'(L)$ can be expressed as an integration involving functions $\mP_{L^2}$. Convergence of this integration is expected, given that $c'(L)$ is generally finite. This expectation implies that $\int_{0}^{\epsilon}d\lambda\mP_{L^2}\lambda^{-1}$ should yield a constant for any positive $\epsilon$, ensuring convergence. Consequently, we obtain $\mP_{L^2}(\epsilon)\epsilon^{-1}\to C$, where $C$ is a constant. Consequently, we derive:
\bea
L^2\lim_{\lambda\to 0} \tilde{\mP}_{L^2}(\lambda) \log\lambda=L^2\lim_{\lambda\to 0} \tilde{\mP}_{L^2}(\lambda)\lambda^{-1} \lambda\log\lambda\to 0.
\eea
The other boundary term $-L^2 \tilde{\mP}_{L^2}(\lambda_m) \log\lambda_m$ is typically non-zero. Since we have $0\le \mP_{L^2}$ and $0<\lambda_m<1$, this term is positive.

The integration part in (\ref{secondfinal}) closely resembles (\ref{firstderivative}) when replacing $\mP_L$ with $L(1+L\frac{\partial }{\partial L})\mP_L$. Similar to our approach in the previous section, the nature of $L(1+L\frac{\partial }{\partial L})\mP_L$ is intricately connected to the sign of the integration result.

\section{Conclusion and discussion}\label{section_conclusion}
 In this paper, we introduce a series of functions designed to characterize the dependence of the entanglement spectrum on parameters. These functions bear resemblance to the density of eigenstate $\mathcal{P}$ extensively discussed in prior literature. Our novel functions, such as $\mathcal{P}{\alpha_J}$ and $\mathcal{P}{\alpha_{J_1}\alpha_{J_2}}$, encapsulate crucial information regarding $\frac{\partial \lambda_i}{\partial \alpha_J}$ and $\frac{\partial^2 \lambda_i}{\partial \alpha_{J_1}\partial \alpha_{J_2}}$. The evaluation of these functions can be accomplished through the utilization of R\'enyi entropy. Notably, functions of the same order exhibit intriguing relationships, e.g., Eq.(\ref{r2}). Furthermore, we demonstrate that these relationships can be derived from their definitions in section.\ref{consistentsection}. However, our study reveals limitations in obtaining all these functions solely through the inverse Laplace transformation method employed in this paper. It appears that alternative methodologies or additional information beyond R\'enyi entropy may be necessary to obtain a complete set of these functions. We will delve into exploring these avenues in the near future.

If we make the additional assumption that the derivative of a given eigenvalue $\lambda_i$ with respect to $\alpha_J$ remains a function of $\lambda_i$, i.e., $\frac{\partial \lambda_i}{\partial \alpha_J}=f(\lambda_i,\alpha_J)$, an intriguing differential equation (\ref{diffequation}) governing $\lambda_i$ can be derived. Solving this differential equation enables the reconstruction of the form of $\lambda_i$. In our examination of a single interval within a vacuum state, we explicitly demonstrate how to derive the dependence of $\lambda_i$ on the subsystem size $L$ using the functions $\mathcal{P}$ and $\mathcal{P}_L$. Remarkably, this outcome aligns with the methodology involving the mapping of the modular Hamiltonian to a cylinder, presenting an interesting application of these functions. However, it's crucial to highlight a significant limitation in this process. The assumption that $\frac{\partial \lambda_i}{\partial \alpha_J}=f(\lambda_i,\alpha_J)$ may not hold universally across all scenarios. Care must be exercised when employing this assumption. An intriguing avenue for exploration involves relaxing this assumption, such as considering whether $\frac{\partial \lambda_i}{\partial \alpha_J}$ depends on all the eigenvalues. Yet, pursuing this route ultimately leads to a series of complex partial differential equations that prove challenging to solve.
Obtaining the exact form of eigenvalues of entanglement Hamiltonian in QFTs remains an extremely challenging problem. While our current findings are constrained, we anticipate that our framework serves as a potential method to reconstruct the eigenvalues of the entanglement Hamiltonian using Renyi entropy.

In several instances, the R\'enyi entropy can be obtained using replica methods. Our paper showcases various examples illustrating how to derive the functions introduced in our study. These instances encompass scenarios such as a single interval in a vacuum state, arbitrary states for a short interval in 2-dimensional CFTs, perturbation states in the general case, and holographic QFTs. Our primary focus lies on understanding the functions $\mathcal{P}$ and $\mathcal{P}_{\alpha_J}$ within these examples.

Calculations in 2-dimensional CFTs yield straightforward results. For perturbation states, we obtain exact expressions at the leading order of the perturbation. The final forms of $\mathcal{P}$ and $\mathcal{P}_{\alpha_J}$ offer insightful explanations. These results become applicable when the R\'enyi entropy of the perturbation state is known.

In the context of holographic theory, a fascinating finding arises for the function $\mathcal{P}{\alpha_J}$. In the semiclassical limit $G \to 0$, where $G$ represents the gravitational constant, the zero point of this function is identified as $\lambda_0 = e^{-S}$ or $t_0 = S - S^{\infty}$. Here, $S$ denotes the Entanglement Entropy (EE), and $S^{\infty}$ signifies the minimal entropy, defined as $S^{\infty} := \lim_{n \to \infty} S^{(n)}$. Intriguingly, the value of $\lambda_0$ or $t_0$ also emerges in the approximated state for $\rho_A$ constructed in \cite{Guo:2020roc}. In that work, the author observes the density of eigenstates approaching a Dirac delta function at the value $t_0 = S - S^{\infty}$. While the relationship between these two findings remains elusive, they signify distinct features of the geometric states in holographic theory. Specifically, the entanglement spectra of these geometric states exhibit peculiar properties near the value $t_0$. Further exploration into this phenomenon is planned for our future investigations.

 The functions introduced in our study are intricately connected to the R\'enyi entropy and its derivatives through Laplace transformations. In principle, they should equate to the R\'enyi entropy since the Laplace transformation is reversible. One might question the necessity of investigating these functions. This parallels the field of signal processing where Fourier or Laplace transformations are employed to convert signals into the dual space. Occasionally, the signal in the dual space offers more intuitive insights. Similarly, while the R\'enyi entropy encapsulates rich information regarding entanglement spectra, the functions we introduced serve as a method to extract this entanglement spectrum information.

The functions associated with $\lambda$ or $t$ can be viewed as the dual space counterparts of the R\'enyi index $n$. Specifically, in the context of holographic theory, these functions have proven useful in comprehending fixed area states and QEC codes for AdS/CFT\cite{Guo:2021tzs}. Particularly, these functions in the dual space are anticipated to hold significant applications in elucidating the properties of geometric states. They offer an alternative perspective to understand and explore the intricacies of entanglement spectra that might not be readily apparent from the R\'enyi entropy alone.

~\\~\\~\\

{\bf Acknowledgements}

WZG is supposed by the National Natural Science Foundation of China under Grant No.12005070 and the Fundamental Research Funds for the Central Universities under Grants NO.2020kfyXJJS041.

\appendix
\section{Details of the calculations for general setup}
\subsection{The proof of general relation (\ref{generally})}\label{prove1}
The proof is straightforward by using the definition and property of delta function, which is shown as follows.
\begin{align}
& \frac{\partial }{\partial \alpha_K} \mP_{(\alpha_{J_{11}}... \alpha_{J_{1m_1}})(\alpha_{J_{21}}... \alpha_{J_{2m_2}})...(\alpha_{J_{n1}}... \alpha_{J_{n m_n}})} \nn \\
=&\frac{\partial }{\partial \alpha_K} \sum_i \frac{\p^{m_1} \lambda_i}{\p \alpha_{J_{11}}...\p {\alpha_{J_{1m_1}}}} \frac{\p^{m_2} \lambda_i}{\p \alpha_{J_{21}}...\p {\alpha_{J_{2m_2}}}} ... \frac{\p^{m_n} \lambda_i}{\p \alpha_{J_{n1}}...\p {\alpha_{J_{n m_n}}}}\delta(\lambda_i -\lambda) \nn \\
=&\sum_i \frac{\p^{m_1+1} \lambda_i}{\p \alpha_{J_{11}}...\p {\alpha_{J_{1m_1}}}\p \alpha_K} \frac{\p^{m_2} \lambda_i}{\p \alpha_{J_{21}}...\p {\alpha_{J_{2m_2}}}} ... \frac{\p^{m_n} \lambda_i}{\p \alpha_{J_{n1}}...\p {\alpha_{J_{n m_n}}}}\delta(\lambda_i -\lambda) \nn \\
&+\sum_i \frac{\p^{m_1} \lambda_i}{\p \alpha_{J_{11}}...\p {\alpha_{J_{1m_1}}}} \frac{\p^{m_2+1} \lambda_i}{\p \alpha_{J_{21}}...\p {\alpha_{J_{2m_2}}}\p \alpha_K} ... \frac{\p^{m_n} \lambda_i}{\p \alpha_{J_{n1}}...\p {\alpha_{J_{n m_n}}}}\delta(\lambda_i -\lambda) \nn \\
&+... \nn \\
&+\sum_i \frac{\p^{m_1} \lambda_i}{\p \alpha_{J_{11}}...\p {\alpha_{J_{1m_1}}}} \frac{\p^{m_2} \lambda_i}{\p \alpha_{J_{21}}...\p {\alpha_{J_{2m_2}}}} ... \frac{\p^{m_n+1} \lambda_i}{\p \alpha_{J_{n1}}...\p {\alpha_{J_{n m_n}}}\p \alpha_K}\delta(\lambda_i -\lambda) \nn \\
&+\sum_i \frac{\p^{m_1} \lambda_i}{\p \alpha_{J_{11}}...\p {\alpha_{J_{1m_1}}}} \frac{\p^{m_2} \lambda_i}{\p \alpha_{J_{21}}...\p {\alpha_{J_{2m_2}}}} ... \frac{\p^{m_n} \lambda_i}{\p \alpha_{J_{n1}}...\p {\alpha_{J_{n m_n}}}}\frac{\partial }{\partial \alpha_K}\delta(\lambda_i -\lambda) \nn \\
=&\sum_i \frac{\p^{m_1+1} \lambda_i}{\p \alpha_{J_{11}}...\p {\alpha_{J_{1m_1}}}\p \alpha_K} \frac{\p^{m_2} \lambda_i}{\p \alpha_{J_{21}}...\p {\alpha_{J_{2m_2}}}} ... \frac{\p^{m_n} \lambda_i}{\p \alpha_{J_{n1}}...\p {\alpha_{J_{n m_n}}}}\delta(\lambda_i -\lambda) \nn \\
&+\sum_i \frac{\p^{m_1} \lambda_i}{\p \alpha_{J_{11}}...\p {\alpha_{J_{1m_1}}}} \frac{\p^{m_2+1} \lambda_i}{\p \alpha_{J_{21}}...\p {\alpha_{J_{2m_2}}}\p \alpha_K} ... \frac{\p^{m_n} \lambda_i}{\p \alpha_{J_{n1}}...\p {\alpha_{J_{n m_n}}}}\delta(\lambda_i -\lambda) \nn \\
&+... \nn \\
&+\sum_i \frac{\p^{m_1} \lambda_i}{\p \alpha_{J_{11}}...\p {\alpha_{J_{1m_1}}}} \frac{\p^{m_2} \lambda_i}{\p \alpha_{J_{21}}...\p {\alpha_{J_{2m_2}}}} ... \frac{\p^{m_n+1} \lambda_i}{\p \alpha_{J_{n1}}...\p {\alpha_{J_{n m_n}}}\p \alpha_K}\delta(\lambda_i -\lambda) \nn \\
&+\sum_i \frac{\p^{m_1} \lambda_i}{\p \alpha_{J_{11}}...\p {\alpha_{J_{1m_1}}}} \frac{\p^{m_2} \lambda_i}{\p \alpha_{J_{21}}...\p {\alpha_{J_{2m_2}}}} ... \frac{\p^{m_n} \lambda_i}{\p \alpha_{J_{n1}}...\p {\alpha_{J_{n m_n}}}}\frac{\partial \lambda_i}{\partial \alpha_K}*[-\frac{\partial}{\partial \lambda}\delta(\lambda_i -\lambda)] \nn \\
=&\mP_{(\alpha_{J_{11}}... \alpha_{J_{1m_1}} \alpha_K)(\alpha_{J_{21}}... \alpha_{J_{2m_2}})...(\alpha_{J_{n1}}... \alpha_{J_{n m_n}})} \nn \\
&+\mP_{(\alpha_{J_{11}}... \alpha_{J_{1m_1}} )(\alpha_{J_{21}}... \alpha_{J_{2m_2}}\alpha_K)...(\alpha_{J_{n1}}... \alpha_{J_{n m_n}})} \nn \\
&+...... \nn \\
&+\mP_{(\alpha_{J_{11}}... \alpha_{J_{1m_1}} )(\alpha_{J_{21}}... \alpha_{J_{2m_2}})...(\alpha_{J_{n1}}... \alpha_{J_{n m_n}}\alpha_K)} \nn \\
&-\frac{\partial }{\partial \lambda} \mP_{(\alpha_{J_{11}}... \alpha_{J_{1m_1}} )(\alpha_{J_{21}}... \alpha_{J_{2m_2}})...(\alpha_{J_{n1}}... \alpha_{J_{n m_n}})(\alpha_K)}.    
\end{align}
\subsection{The formula derivation of (\ref{higher_times}) $\mP_{\alpha_J^m}$ and (\ref{higher_derivative}) $\mP_{\alpha_{J m}}$}\label{prove2}
The derivation of $\mP_{\alpha_J^m}$ is trivial, with our assumption
 $\frac{\p \lambda_i}{\p \alpha_J}=f(\lambda_i,\alpha_J)$
 \begin{align}
\mP_{\alpha_J^m}(\lambda)&=\sum_i (\frac{\p \lambda_i}{\p \alpha_J})^m \delta(\lambda_i-\lambda)\nn\\
&=f(\lambda,\alpha_J)^m\mP(\lambda).
 \end{align}
Since the case where $m=2$ for $\mP_{\alpha_{J m}}$ we've already deduced, the case of $m>2$ can be proved by mathematical induction. Since we assume that the formula holds for $m-1$, we have
\begin{align}
\mP_{\alpha_{J m-1}}(\lambda)&=\frac{D^{m-2}f(\lambda,\alpha_J)}{D \alpha_J^{m-2}}\mP(\lambda)\nn\\
\mP_{\alpha_{J m-1}}(\lambda)&=g(\lambda,\alpha_J)\mP(\lambda)\nn\\
\sum_i\frac{d^{m-1}}{d \alpha_J^{m-1}}f(\lambda_i,\alpha_J) \delta(\lambda_i-\lambda)&=\sum_ig(\lambda_i,\alpha_J) \delta(\lambda_i-\lambda)\nn\\
\frac{d^{m-1}}{d \alpha_J^{m-1}}f(\lambda_i,\alpha_J) &=g(\lambda_i,\alpha_J),
\end{align}
where we define
\begin{align}
g(\lambda,\alpha_J):=\frac{D^{m-2}f(\lambda,\alpha_J)}{D \alpha_J^{m-2}}.
\end{align}
So in the case of $m$, we have
\begin{align}
\mP_{\alpha_{J m}}(\lambda)&=\sum_i\frac{d}{d \alpha_J}\frac{d^{m-1}}{d \alpha_J^{m-1}}f(\lambda_i,\alpha_J) \delta(\lambda_i-\lambda)\nn\\
\mP_{\alpha_{J m}}(\lambda)&=\sum_i\frac{d}{d \alpha_J}g(\lambda_i,\alpha_J) \delta(\lambda_i-\lambda)\nn\\
&=\sum_i (\frac{\partial g(\lambda_i,\alpha_J)}{\partial \lambda_i} \frac{\partial \lambda_i}{\partial \alpha_J} + \frac{\partial g(\lambda_i,\alpha_J)}{\partial \alpha_J})\delta(\lambda_i-\lambda)\nn\\
&=\sum_i (\frac{\partial g(\lambda_i,\alpha_J)}{\partial \lambda_i} f(\lambda_i,\alpha_J) +\frac{\partial g(\lambda_i,\alpha_J)}{\partial \alpha_J})\delta(\lambda_i-\lambda)\nn\\
&=(\frac{\partial g(\lambda,\alpha_J)}{\partial \lambda} f(\lambda,\alpha_J) + \frac{\partial g(\lambda,\alpha_J)}{\partial \alpha_J} )\sum_i \delta(\lambda_i-\lambda)\nn\\
&=\frac{D g(\lambda,\alpha_J)}{D\alpha_J}\mP(\lambda)\nn\\
&=\frac{D^{m-1}f(\lambda,\alpha_J)}{D \alpha_J^{m-1}}\mP(\lambda).
\end{align}

\section{Consistent check of the functions for vacuum state}\label{sectionconsistent}
In the main text we obtain the function $\mP$ and $\mP_l$ for the one interval in the vacuum state of CFTs. In this section we would like to check the consistent relation (\ref{consistent1}). Since $t=-b-\log\lambda$, we have $\frac{\p t}{\p L}=-\frac{\p b}{\p L}$, so
\begin{align}
&\frac{\partial}{\partial L}  P(\lambda)=\frac{1}{\lambda} \frac{\partial}{\partial L} (\frac{\sqrt{b} I_1\left(2 \sqrt{b t}\right)}{\sqrt{t}}+\delta (t))\nn\\
&\phantom{\frac{\partial}{\partial L}  P(\lambda)}=\frac{1}{\lambda} [\frac{\partial}{\partial b} (\frac{\sqrt{b} I_1\left(2 \sqrt{b t}\right)}{\sqrt{t}})\frac{\partial b}{\partial L}+\frac{\partial}{\partial t} (\frac{\sqrt{b} I_1\left(2 \sqrt{b t}\right)}{\sqrt{t}})\frac{\partial t}{\partial L}]+\frac{1}{\lambda} \frac{\partial t}{\partial L} \delta' (t)\nn\\
&\phantom{\frac{\partial}{\partial L}  P(\lambda)}=\frac{1}{\lambda} [\frac{c}{6L} (\frac{I_1\left(2 \sqrt{b t}\right)}{2 \sqrt{b t}}+\frac{1}{2} \left(I_0\left(2 \sqrt{b t}\right)+I_2\left(2 \sqrt{b t}\right)\right))\nn\\
&\phantom{\frac{\partial}{\partial L}  P(\lambda)=}-\frac{c}{6L}(\frac{b \left(I_0\left(2 \sqrt{b t}\right)+I_2\left(2 \sqrt{b t}\right)\right)}{2 t}-\frac{\sqrt{b} I_1\left(2 \sqrt{b t}\right)}{2 t^{3/2}})]-\frac{c}{6L} \frac{1}{\lambda} \delta' (t)\nn\\
&\phantom{\frac{\partial}{\partial L}  P(\lambda)}=\frac{1}{\lambda} \frac{c}{6L} \frac{I_1\left(2 \sqrt{b t}\right)}{2 \sqrt{b t}}+\frac{1}{\lambda} \frac{c}{6L} \frac{\sqrt{b} I_1\left(2 \sqrt{b t}\right)}{2 t^{3/2}})\nn \\
&\phantom{\frac{\partial}{\partial L}  P(\lambda)=}+\frac{1}{\lambda} \frac{c}{6L} \frac{1}{2} \frac{t-b}{t} \left(I_0\left(2 \sqrt{b t}\right)+I_2\left(2 \sqrt{b t}\right)\right)-\frac{1}{\lambda} \frac{c}{6L}\delta' (t)    
\end{align}
\begin{align}
-\frac{\partial}{\partial \lambda}  P_L(\lambda)&=\frac{\partial}{\partial \lambda} \times \frac{c}{6L}(\frac{(b-t) I_1\left(2 \sqrt{b t}\right)}{\sqrt{b t}}+\delta (t))\nn\\
&=\frac{c}{6L}\frac{\partial}{\partial t}(\frac{(b-t) I_1\left(2 \sqrt{b t}\right)}{\sqrt{b t}})\frac{\partial t}{\partial \lambda}+ \frac{c}{6L} \frac{\partial t}{\partial \lambda} \delta' (t)\nn\\
&=\frac{1}{\lambda} \frac{c}{6L} \frac{I_1\left(2 \sqrt{b t}\right)}{2 \sqrt{b t}}+\frac{1}{\lambda} \frac{c}{6L} \frac{\sqrt{b} I_1\left(2 \sqrt{b t}\right)}{2 t^{3/2}})\nn\\
&\quad+\frac{1}{\lambda} \frac{c}{6L} \frac{1}{2} \frac{t-b}{t} \left(I_0\left(2 \sqrt{b t}\right)+I_2\left(2 \sqrt{b t}\right)\right)-\frac{1}{\lambda} \frac{c}{6L}\delta' (t)\nn\\
&=left.    
\end{align}
On the other hand, since we have (\ref{spectra1}), let's rewrite
\begin{align}
&\mP_L(\lambda)=-\frac{c}{6L}\left[\frac{(b-t) I_1\left(2 \sqrt{bt}\right)}{\sqrt{b t}}+\delta (t)\right]=-\frac{c}{6L} \mP_{\alpha_J}(\lambda),\nn \\
&\mP_c(\lambda)=-\frac{\log{L}}{6}\left[\frac{(b-t) I_1\left(2 \sqrt{bt}\right)}{\sqrt{b t}}+\delta (t)\right]=-\frac{\log{L}}{6}\mP_{\alpha_J}(\lambda),
\end{align}
so, we have
\begin{align}
\frac{\p}{\p c}\mP_L(\lambda)&=-\frac{1}{6L} \mP_{\alpha_J}(\lambda)-\frac{c}{6L}\frac{\log{L}}{6}\frac{\p}{\p b}\mP_{\alpha_J}(\lambda)\nn\\
\frac{\p}{\p L}\mP_c(\lambda)&=-\frac{1}{6L} \mP_{\alpha_J}(\lambda)-\frac{\log{L}}{6}\frac{c}{6L}\frac{\p}{\p b}\mP_{\alpha_J}(\lambda)=\frac{\p}{\p c}\mP_L(\lambda)
\end{align}
\section{Reconstruction of the eigenvalue}
In section.\ref{reconstructionsection} we use further assumption that $\frac{\p \lambda_i}{\p L}=f(\lambda_i)$ and the functions $\mP$ and $\mP_L$ to reconstruct the eigenvalues of $\rho_A$ . In principle one could also choose other parameters, such as $c$. But one would obtain the wrong results as we will show below. 
Using (\ref{spectra1}) and the assumption $\frac{\p \lambda_i}{\p c}=f(\lambda_i)$ we have
\bea
\frac{\partial \lambda_i}{\partial c}=\frac{-\log{\lambda_i}-2b}{b}\frac{\log{L}}{6}\lambda_i,
\eea
which can be solved as
\bea\label{lamda2}
\lambda_i=e^{-\frac{\tilde{C}_i}{c}-b},
\eea
where $\tilde{C}_i$ are constants unrelated to $c$. This is inconsistent with the form $e^{-\frac{\Delta_i -\frac{c}{24}}{W}-b}$ that is derived in section.\ref{sectionconformalmap}.  By the normalization of $\rho_A$ we have
\bea
\sum_i e^{-\frac{\Delta_i-\frac{c}{24}}{W}-b}=1.
\eea
Using this one could find that $\frac{\p \Delta_i}{\p c}$ should satisfy the constraint
\bea
\sum_i \frac{\p \Delta_i}{\p c} e^{-\frac{\Delta_i}{W}}=-W e^{b-\frac{c}{24W}}\left(\frac{\p b}{\p c} -\frac{1}{24W}\right),
\eea
which means $\frac{\p \Delta_i}{\p c}$ is not only a function of $\Delta_i$ but depends on other $\Delta_j$ ($j\ne i$). Therefore, the assumption $\frac{\p \lambda_i}{\p c}=f(\lambda_i)$ is incorrect. 
\section{Details of the calculations for short interval}\label{shortsectionapp}
For the  thermal state, we have 
\bea
\langle T \rangle_\beta=\langle \bar T \rangle_\beta=-\frac{\pi^2c}{6\beta^2}, \langle A \rangle_\beta=\langle \overline{A} \rangle\beta=\frac{\pi^4c(5c+22)}{180\beta^4}. 
\eea
Thus, we obtain
\bea
k_2=\frac{\pi^2c}{36\beta^2}, k_4=-\frac{11\pi^4c}{12960\beta^4}, k'_4&=-\frac{\pi^4c}{12960\beta^4}.
\eea
Taking the above results into (\ref{P4}) we have
\begin{align}
\mP&=\lambda^{-1}[\frac{\sqrt{b_0}}{\sqrt{t}}I_1+\delta(t)+(k_2l^2+(2k_4-10k'_4)l^4)I_0+\frac{1}{2}k_2^2l^4\frac{\sqrt{t}}{\sqrt{b_0}}I_1\nn\\
&=\lambda^{-1}[\frac{\sqrt{b_0}}{\sqrt{t}}I_1+\delta(t)+(\frac{\pi^2c}{36\beta^2}l^2-\frac{\pi^4c}{1080\beta^4}l^4)I_0+\frac{\pi^4c^2}{2596\beta^4}l^4\frac{\sqrt{t}}{\sqrt{b_0}}I_1,
\end{align}
where the argument of $I_n$ is $2\sqrt{b_0t}$, same as below.
The R\'enyi entropy of the thermal state is given by
\bea
S^{(n)}=\frac{c}{6}(1+\frac{1}{n})\log[\frac{\beta}{\pi\epsilon}\sinh(\frac{\pi l}{\beta})]
\eea
so we can get $\mP$ by 
\begin{align}
\mP'&=\lambda^{-1} \mL^{-1}\left[e^{nb_T+(1-n)S^{(n)}_T}\right]\nn\\
&=\frac{1}{\lambda}(\frac{\sqrt{b_T} I_1\left(2 \sqrt{b_T t}\right)}{\sqrt{t}}+\delta (t))
\end{align}
with
\bea
b_T=\frac{c}{6}\log[\frac{\beta}{\pi\epsilon}\sinh(\frac{\pi l}{\beta})].
\eea
Expanding $\mP'$ upto the order $(\frac{l}{\beta})^4$, we find
\begin{align}
\mP'&=\lambda^{-1}[\frac{\sqrt{b_0}}{\sqrt{t}}I_1+\delta(t)+\frac{c\pi^2(I_1+\sqrt{b_0 t}I_2)l^2}{36\beta^2\sqrt{b_0 t}}\nn\\
&+\frac{[5\pi^4c^2tI_1-12\pi^4c(I_1+\sqrt{b_0 t}I_2)]l^4}{12960\beta^4\sqrt{b_0 t}}]+O(l^5).
\end{align}
By using the relation $I_{n-1}(x)-I_{n+1}(x)=\frac{2nI_n(x)}{x}$, we have
\begin{align}
I_0(2\sqrt{b_0 t})-I_2(2\sqrt{b_0t})&=\frac{2I_1(2\sqrt{b_0t})}{2\sqrt{b_0t}}\nn\\
I_1(2\sqrt{b_0t})+\sqrt{b_0t}I_2(2\sqrt{b_0t})&=\sqrt{b_0t}I_0(2\sqrt{b_0t}).
\end{align}
Thus we find
\begin{align}
\mP'&=\lambda^{-1}[\frac{\sqrt{b_0}}{\sqrt{t}}I_1+\delta(t)+(\frac{\pi^2c}{36\beta^2}l^2-\frac{\pi^4c}{1080\beta^4}l^4)I_0+\frac{\pi^4c^2}{2596\beta^4}l^4\frac{\sqrt{t}}{\sqrt{b_0}}I_1\nn\\
&=\mP.
\end{align}
One could also check the results for $\mP_l$ by same method.
\section{Consistent check of the functions for perturbation states}\label{Consistentcheck_perturbation}
Before checking the relation (\ref{r1}) of (\ref{P_perturbation}) and (\ref{PaJ_perturbation}), we want to find the relation between $\mP_{\delta}$ and $\mP_{(\delta \alpha_J)}$ and $\mP_{(\delta)( \alpha_J)}$ first. By the definitions, we have
\begin{align}\label{r2}
\frac{\p}{\p \alpha_J}\mP_{\delta}(\lambda)&=\frac{\p}{\p \alpha_J}\sum_i \delta \lambda_i\delta(\lambda^0_i-\lambda)\nn\\
&=\sum_i \frac{\p \delta \lambda_i}{\p \alpha_J}\delta(\lambda^0_i-\lambda)+\sum_i \delta \lambda_i\frac{\p \lambda^0_i}{\p \alpha_J}\delta'(\lambda^0_i-\lambda)\nn\\
&=\mP_{(\delta \alpha_J)}(\lambda)-\frac{\p}{\p \lambda}\mP_{(\delta)( \alpha_J)}(\lambda)
\end{align}
By using (\ref{P_perturbation}) and (\ref{PaJ_perturbation}) and (\ref{r2})
\begin{align}
&\frac{\p}{\p \alpha_J}\mP(\lambda)\nn\\
=&\frac{\p}{\p \alpha_J}[\mP_0(\frac{\lambda^0_m}{\lambda_m}\lambda)-\delta b\lambda\frac{\p\mP_0}{\p \lambda}(\frac{\lambda^0_m}{\lambda_m}\lambda)-\frac{\p\mP_{\delta}}{\p \lambda}(\frac{\lambda^0_m}{\lambda_m}\lambda)]\nn\\
=&\frac{\p\mP_0}{\p \alpha_J}(\frac{\lambda^0_m}{\lambda_m}\lambda)+\frac{\p\mP_0(\frac{\lambda^0_m}{\lambda_m}\lambda)}{\p\frac{\lambda^0_m}{\lambda_m}\lambda}\frac{\p\frac{\lambda^0_m}{\lambda_m}\lambda}{\p \alpha_J}-\frac{\p\delta b}{\p \alpha_J}\lambda\frac{\p\mP_0}{\p \lambda}(\frac{\lambda^0_m}{\lambda_m}\lambda)\nn\\
&-\delta b\lambda\frac{\p^2\mP_0}{\p \lambda\p\alpha_J}(\frac{\lambda^0_m}{\lambda_m}\lambda)+O(\delta \rho^2)-\frac{\p^2\mP_{\delta}}{\p \lambda\p\alpha_J}(\frac{\lambda^0_m}{\lambda_m}\lambda)+O(\delta \rho^2)\nn\\
=&-\frac{\p\mP^0_{\alpha_J}}{\p \lambda}(\frac{\lambda^0_m}{\lambda_m}\lambda)+\delta b\lambda\frac{\p^2\mP^0_{\alpha_J}}{\p \lambda^2}(\frac{\lambda^0_m}{\lambda_m}\lambda)-[\frac{\p\mP_{(\delta\alpha_J)}}{\p \lambda}-\frac{\p^2\mP_{(\delta)(\alpha_J)}}{\p \lambda^2}](\frac{\lambda^0_m}{\lambda_m}\lambda)
\end{align}
and
\begin{align}
&-\frac{\p\mP_{\alpha_J}}{\p \lambda}(\lambda)\nn\\
=&-\frac{\p}{\p \lambda}[\mP^0_{\alpha_J}(\frac{\lambda^0_m}{\lambda_m}\lambda)-\delta b\lambda\frac{\p\mP_{\alpha_J}^0}{\p\lambda}(\frac{\lambda^0_m}{\lambda_m}\lambda)+\mP_{(\delta \alpha_J)}(\frac{\lambda^0_m}{\lambda_m}\lambda)-\frac{\p\mP_{(\delta)(\alpha_J)}}{\p\lambda}(\frac{\lambda^0_m}{\lambda_m}\lambda)]\nn\\
=&-(1+\delta b)\frac{\p \mP^0_{\alpha_J}}{\p \lambda}(\frac{\lambda^0_m}{\lambda_m}\lambda)+\delta b\frac{\p\mP_{\alpha_J}^0}{\p\lambda}(\frac{\lambda^0_m}{\lambda_m}\lambda)+\delta b(1+\delta b)\lambda\frac{\p^2\mP_{\alpha_J}^0}{\p\lambda^2}(\frac{\lambda^0_m}{\lambda_m}\lambda)\nn\\
&-[(1+\delta b)\frac{\p\mP_{(\delta\alpha_J)}}{\p \lambda}(\frac{\lambda^0_m}{\lambda_m}\lambda)-(1+\delta b)\frac{\p^2\mP_{(\delta)(\alpha_J)}}{\p \lambda^2}(\frac{\lambda^0_m}{\lambda_m}\lambda)]\nn\\
=&-\frac{\p\mP^0_{\alpha_J}}{\p \lambda}(\frac{\lambda^0_m}{\lambda_m}\lambda)+\delta b\lambda\frac{\p^2\mP^0_{\alpha_J}}{\p \lambda^2}(\frac{\lambda^0_m}{\lambda_m}\lambda)-[\frac{\p\mP_{(\delta\alpha_J)}}{\p \lambda}-\frac{\p^2\mP_{(\delta)(\alpha_J)}}{\p \lambda^2}](\frac{\lambda^0_m}{\lambda_m}\lambda)+O(\delta\rho^2)\nn\\
=&\frac{\p}{\p \alpha_J}\mP(\lambda)
\end{align}

\end{document}